\documentclass[10pt,a4paper]{article}
\usepackage[utf8]{inputenc}
\usepackage{amsmath,a4wide}
\usepackage{amsfonts}
\usepackage{amssymb}
\usepackage{siunitx}
\usepackage{booktabs}
\usepackage{graphicx}
\usepackage{physics}
\usepackage{multicol}
\usepackage{mathtools}
\usepackage[position=b]{subcaption}
\usepackage[space]{grffile}
\usepackage{float,hyperref}
\usepackage{bbm}
\usepackage{wrapfig,parskip}
\usepackage[toc,page]{appendix}
\usepackage{cancel,a4wide}
\usepackage[affil-it]{authblk}

\newcommand{\I}{\mathbbm{1}}
\newcommand{\Lp}{\mathcal{L}}
\newcommand{\p}{\partial}
\newcommand{\F}{\mathcal{F}}
\newcommand{\Nres}{N_{\text{res}}}

\title{\textbf{Spectral representation of lattice gluon and ghost propagators at zero temperature}}
\author{David Dudal$^{a, b}$\thanks{\href{mailto:david.dudal@kuleuven.be}{david.dudal@kuleuven.be}}, Orlando Oliveira$^c$\thanks{\href{mailto:orlando@uc.pt}{orlando@uc.pt}}, Martin Roelfs$^{a}$\thanks{\href{mailto:martin.roelfs@kuleuven.be}{martin.roelfs@kuleuven.be}; Corresponding author}, Paulo Silva$^c$\thanks{\href{mailto:psilva@uc.pt}{psilva@uc.pt}}}
\affil{\footnotesize $^{a}$ KU Leuven Campus Kortrijk--Kulak, Department of Physics, Etienne Sabbelaan 53 bus 7657, 8500 Kortrijk, Belgium\\
	$^{b}$ Ghent University, Department of Physics and Astronomy, Krijgslaan 281-S9, 9000 Gent, Belgium\\
	$^{c}$ Centro de F\'{i}sica Computacional, Departamento de F\'{i}sica, Universidade de Coimbra, 3004-516 Coimbra, Portugal}

\begin{document}
	\maketitle
	\begin{abstract}
		We consider the analytic continuation of Euclidean propagator data obtained from 4D simulations to Minkowski space.
		In order to perform this continuation, the common approach is to first extract the Källén-Lehmann spectral density of the field. Once this is known, it can be extended to Minkowski space to yield the Minkowski propagator. However, obtaining the Källén-Lehmann spectral density from propagator data is a well known ill-posed numerical problem. To regularize this problem we implement an appropriate version of Tikhonov regularization supplemented with the Morozov discrepancy principle. We will then apply this to various toy model data to demonstrate the conditions of validity for this method, and finally to zero temperature gluon and ghost lattice QCD data. We carefully explain how to deal with the IR singularity of the massless ghost propagator. We also uncover the numerically different performance when using two ---mathematically equivalent--- versions of the Källén-Lehmann spectral integral.
	\end{abstract}

	\section{Introduction}

	A plethora of practical Quantum Field Theory calculational tools, both analytical and numerical, have been developed in a Euclidean setting, despite living in a Minkowski spacetime.
	In particular, non-perturbative approaches to Quantum Field Theory tend to rely solely on the Euclidean formulation of the theory due to its technical advantages.
	This is also the case for the discretized lattice approach to Yang-Mills theory, or most of its continuum studies using the Dyson-Schwinger equations, although there are also examples of solving the Dyson-Schwinger equations directly in Minkowski space \cite{Sauli:2002tk, Frederico:2019noo, Solis:2019fzm}.
	The Euclidean formulation accesses only space-like momenta and, therefore, phenomena that are associated with time-like momenta cannot be investigated within Euclidean Quantum Field Theory.
	Moreover, whereas in perturbation theory the analytical continuation of the Euclidean correlation functions into the entire complex momenta Argand plane relies on the usual Wick rotation, it is not clear that the same rule can be applied for the non-perturbative regime.
	Oftentimes the analytical structure of the perturbative correlation functions are known, allowing to go from the Minkowski to Euclidean spacetime and vice-versa in an unambiguous way.
	However, for the non-perturbative regime, the analytical structure of the correlation functions is difficult to determine \textit{per se} and the analytical continuation of the Euclidean space correlation functions into the complex plane becomes a much harder problem.
	Additionally, with numerical data in particular, we only have a discrete set of data at our disposal which we wish to analytically continue to the complex plane, while it is well-known that such a continuation is only unique when departing from a function known over an open subset of $\mathbb{C}$.

	For two-point correlation functions, a possible strategy is to compute the Källén-Lehmann spectral density $\rho(\omega)$ from the Euclidean data and reintegrate $\rho(\omega)$ to determine the propagator for Minkowski $p^2 \leq 0$.
	The spectral density encodes information about the spectra and, at finite temperature, it can be related to transport coefficients and thermodynamical properties of e.g.~the quark-gluon plasma \cite{Meyer:2011gj}.
	In this paper we discuss the computation of the spectral density from Euclidean data, focusing on the inversion of the propagator data and on how to improve the method set out in \cite{Dudal:2013yva}.

	One might wonder why it is relevant to investigate the spectral representations of gauge variant degrees of freedom like gluons, ghosts or quarks in the first place, since in general their propagators will be gauge dependent and therefore so will the corresponding spectral densities. But because the physical bound state spectrum is gauge invariant despite being constructed from gauge variant gluon, ghost and quark propagators \cite{Alkofer:2000wg}, the spectral function of a bound state propagator must therefore be nontrivially influenced by the analytic structure of the underlying constituents, see for instance \cite{Alkofer:2000wg,Roberts:1994dr,Maris:2003vk,Bhagwat:2002tx,Windisch:2012sz,Eichmann:2016yit} for discussions and examples. This explains the relevance of various studies devoted to the spectral properties of a priori unphysical degrees of freedom. The Landau gauge in particular is a standard choice to study the physical bound state spectrum of QCD, see \cite{Alkofer:2000wg,Roberts:1994dr,Maris:2003vk,Bhagwat:2002tx,Eichmann:2016yit}, exactly because there are ample estimates, both from the continuum as from the lattice side, about the necessary Green functions that serve as input for the Bethe--Salpeter equations. A beautiful example is the seminal paper \cite{Sanchis-Alepuz:2015hma}, where the Bethe-Salpeter equations for glueballs are considered, making explicit use of the spectral properties encoded in the spectral functions of gluons and ghosts, reported earlier in \cite{Strauss:2012dg}. Our research output gives a direct lattice verification of how trustworthy these latter gluon and ghost spectral functions actually are.

	The story continues at finite temperature, as at some sufficiently high temperature above deconfinement, we expect to achieve some kind of physical quasi-particle behaviour for the spectral functions of deconfined quarks and gluons \cite{Maas:2011se, Oliveira:2019erx}.  From the previous perspective, it may also be interesting to note that recent lattice studies went beyond Landau gauge, and saw only a very mild dependence on the gauge parameter for at least gluon and ghost propagators \cite{Cucchieri:2009kk,Bicudo:2015rma,Cucchieri:2018doy}. This suggests that the gluon and ghost spectral densities will themselves only mildly depend on the gauge parameter, so a study of the lattice Landau gauge spectral functions can teach us some generic properties of confined degrees of freedom, whilst allowing to confront other analytical estimates which are based on certain essential underlying assumptions and approximations which do not plague lattice computations.

	The Euclidean two-point function $G(p_4)$ can be expressed in terms of its Källén-Lehmann spectral density $\rho(\omega)$ as (see e.g.~\cite{Negele:1988vy,Peskin:1995ev,Laine:2016hma})
	\begin{align}
		G(p_4) &= \int_{\omega_0}^\infty \frac{2 \omega \rho(\omega)}{p_4^2 + \omega^2} \dd{\omega} \label{eq:kl_rep_p2} \\
		&\equiv  \int_{\omega_0^2}^\infty \frac{\tilde\rho(\mu)}{p_4^2 + \mu} \dd{\mu} \label{eq:kl_rep_p2_mu}
	\end{align}
	with $p_4$ the imaginary frequency and $\omega_0$ an IR-cutoff, potentially zero.
	Henceforth, the representation in Eq.~\eqref{eq:kl_rep_p2_mu} will be referred to as the $p^2$-formalism, the strategy followed earlier in \cite{Dudal:2013yva}.

	From the antisymmetry property for a spectral density corresponding to a propagator of scalar degrees of freedom \cite{Laine:2016hma}, $\rho(-\omega) = - \rho(\omega)$, it follows that Eq.~\eqref{eq:kl_rep_p2} can equivalently be written as
	\begin{align}
		G(p_4) &= \int_{-\infty}^\infty \Omega(\omega, \omega_0) \frac{\rho(\omega)}{\omega - i p_4} \dd{\omega}, \qquad \Omega(\omega, \omega_0) = \begin{cases}
			0	& \abs{\omega} < \omega_0 \\
			1 & \text{otherwise}
		\end{cases}.
		\label{eq:kl_rep_ip}
	\end{align}
	This representation will be referred to as the $ip$-formalism.
	When presented with a finite data set for $G(p_4)$, one can rewrite Eq.~\eqref{eq:kl_rep_p2_mu} or \eqref{eq:kl_rep_ip} as a matrix equation
	\begin{equation}
		\boldsymbol{G} = \boldsymbol{K} \boldsymbol{\rho},
		\label{eq:GKrho}
	\end{equation}
	where $\boldsymbol{K}$ is the respective integral kernel represented as a matrix, and try to solve it for $\rho(\omega)$. Writing $\boldsymbol{K}$ in terms of its singular value decomposition $\sum_{ij} u_i S_{ij} v_j^\dagger$, where $S_{ij}$ is a rectangular diagonal matrix, the least-squares solution to Eq.~\eqref{eq:GKrho} is then given by
	\begin{equation}
		\rho = \sum_{i=1}^{N} \frac{u_i^\dagger G}{s_i}v_i,
		\label{eq:svd_rss}
	\end{equation}
	with $s_i$ being the singular values of the matrix $\boldsymbol{K}$.
	However, since the singular values of the matrix $\boldsymbol{K}$ span a very large range, the matrix is ill-conditioned, meaning that even a small error in the input values can cause a huge variation in the output.          	Therefore, the inversion problem represented in Eq.~\eqref{eq:GKrho} has to be regularized in  order to be able to compute a solution for $\rho(\omega)$.
	A modification has to be made such that the condition number is reduced and the resulting problems limited.

	Tikhonov regularization solves this inversion problem by adding a term proportional to the norm of the solution to the residual sum of squares\footnote{More general regularization strategies with different norms also exist.}:
	\begin{equation}
		J_\alpha = \norm{\boldsymbol{K} \boldsymbol{\rho} - \boldsymbol{G}}_2^2 + \alpha^2 \norm{\boldsymbol{\rho}}_2^2.
		\label{Eq:Tikhonov1}
	\end{equation}
	In terms of the singular values of $\boldsymbol{K}$ the solution is then given by
	\begin{equation}
		\rho = \sum_{i=1}^{N} \frac{s_i^2}{s_i^2 + \alpha^2} \frac{u_i^\dagger G}{s_i} v_i.
		\label{eq:svd_tikhonov}
	\end{equation}
	By comparing with Eq.~\eqref{eq:svd_rss}, we find that the Tikhonov parameter $\alpha^2$ dampens the effect of the smallest singular values $s_i$.

	For Tikhonov regularization to work, an appropriate choice for the parameter $\alpha^2$ has to be made, but there is no unique way to select the value of $\alpha^2$.
	In this paper we will use the Morozov discrepancy principle \cite{Kirsch:1996:IMT:236740}, which seems a good choice to invert lattice data since such data always feature a statistical error $\sigma_i$ for every data point $G_i$. The Morozov discrepancy principle states that $\alpha^2$ should be selected such that
	\begin{equation}
		\norm{\boldsymbol{K} \boldsymbol{\rho} - \boldsymbol{G}}_2^2 = \sum_{i} \sigma_i^2,
		\label{eq:morozov}
	\end{equation}
	where $\sum_{i} \sigma_i^2$ is the total variance in the data.
	The $\alpha^2$ obeying this constraint is guaranteed to be unique \cite{Kirsch:1996:IMT:236740} and this choice for the regularization parameter means that the quality of the reconstruction matches the quality of the original data set. It can be shown that in the limit $\sigma_i \to 0$ and $N \to \infty$, the Morozov solution converges to the exact solution \cite{Kirsch:1996:IMT:236740}.
	It is also worth noting that Tikhonov regularization using the Morozov criterion can alternatively be understood from a Bayesian approach as ``historic MEM'', with a default model $m=0$.
	This default model choice is well motivated, as the UV asymptotics of the ghost and gluon spectral functions predict that they will tend to zero, corresponding with a default model $m=0$ in the UV.
	But the Tikhonov functional Eq. \eqref{Eq:Tikhonov1} is special among different possible choices of prior distributions, as it can be solved analytically and thus numerical difficulties associated with solving non-Gaussian priors can be avoided.

	This paper reports on applying the procedure outlined above to gluon and ghost two-point functions obtained from lattice QCD.
	However, before applying the Tikhonov procedure to such lattice data, two different implementations of Tikhonov regularisation are outlined in Section~\ref{sec:methods}, corresponding to solving either Eq.~\eqref{eq:kl_rep_p2_mu} or Eq.~\eqref{eq:kl_rep_ip}.
	Then, both of these methods were applied to three different toy models, the results of which are detailed in Section~\ref{sec:toy_model_results}.
	The first of these toy spectral densities is an everywhere positive distribution, chosen to represent an observable physical particle.
	Indeed, via the optical theorem the spectral function can be related to an observable probability \cite{Peskin:1995ev}, implying its positive definiteness.
	By contrast, the second and third toy spectral densities display positivity violations, mimicking the expected behaviour for unphysical (confined) particles such as gluons and ghosts.
	Lastly, the two methods were applied to lattice gluon and ghost $T=0$ data sets, the results of which are given in Section~\ref{sec:gluon_and_ghost}.

	\section{Survey of the method}
	\label{sec:methods}

	The lattice data for the propagator come with known statistical uncertainties and, furthermore, given that the different momenta are computed from the same set of gauge configurations, the different momenta are statistically correlated.
	These two effects can be taken into account in the variational principle behind the Tikhonov regularization
	scheme, replacing (\ref{Eq:Tikhonov1}) by the new minimizing functional\footnote{Given that the propagator and the spectral density are real functions of their real argument (as for our data), in
		the minimizing functional one only needs to consider the transpose. For complex valued quantities, in $J_\alpha$ one should consider the Hermitian conjugate rather than the transpose.}
	\begin{equation}
		J_\alpha = \pqty{\boldsymbol{K} \boldsymbol{\rho} - \boldsymbol{G}}^T \boldsymbol{\Sigma}^{-1} \pqty{\boldsymbol{K} \boldsymbol{\rho} - \boldsymbol{G}} + \alpha^2 \boldsymbol{\rho}^T \, \boldsymbol{\rho},
		\label{eq:tikhonov_cov}
	\end{equation}
	where $\boldsymbol{\Sigma}$ is the covariance matrix.
	For lattice data, the covariance matrix can be computed from the different gauge configurations as
	\begin{equation}
		\Sigma(p_i,p_j) = \frac{1}{N_{\mbox{\begin{tiny}Conf\end{tiny}}}} \,
		\sum^{N_{\text{Conf}}}_{k = 1} \Big{(} G_k(p_i) - \expval{G(p_i)} \Big{)} ~\Big{(} G_k(p_j) - \expval{G(p_j)} \Big{)} \ ,
	\end{equation}
	where $N_{\mbox{\begin{tiny}Conf\end{tiny}}}$ is the number of gauge configurations used to compute the propagator,
	$G_k(p_i)$ is the propagator for gauge configuration $k$ at momentum $p_i$, and $\expval{G(p_j)}$ is the lattice estimation for the propagator.
	However, for both the gluon and ghost lattice data we have checked that the covariant matrix is an almost diagonal matrix and, therefore, herein we shall only consider a diagonal covariance matrix $\Sigma_{ij} = \sigma^2_i \delta_{ij}$ (no sum), where $\sigma^2_i$ is the variance of $G(p_i)$:
	\begin{equation}
		\sigma^2(p_i) = \frac{1}{N_{\mbox{\begin{tiny}Conf\end{tiny}}}} \,
		\sum^{N_{\text{Conf}}}_{k = 1} \Big{(} G_k(p_i) - \expval{G(p_i)} \Big{)} ~\Big{(} G_k(p_i) - \expval{G(p_i)} \Big{)} \ .
	\end{equation}
	Solving for the minimum of function (\ref{eq:tikhonov_cov}) involves computing $\pdv*{J_\alpha}{\boldsymbol{\rho}} = 0$, which results in
	\begin{equation}
		\frac{1}{2} \,\pdv{J_\alpha}{\boldsymbol{\rho}} = \boldsymbol{K}^T \boldsymbol{\Sigma}^{-1} \pqty{\boldsymbol{K} \boldsymbol{\rho} - \boldsymbol{G}}
		+ \alpha^2 \boldsymbol{\rho} = 0.
		\label{eq:tikhonov_vec_derivative}
	\end{equation}
	Defining $\boldsymbol{c} \coloneqq \boldsymbol{K} \boldsymbol{\rho} - \boldsymbol{G}$, then gives
	\begin{equation}
		\boldsymbol{\rho} = - \frac{1}{\alpha^2} ~ \boldsymbol{K}^T ~ \boldsymbol{\Sigma}^{-1}  ~ \boldsymbol{c} \ .
		\label{eq:tikhonov_vec_rho}
	\end{equation}
	However, since $\boldsymbol{c}$ itself depends on $\boldsymbol{\rho}$, substituting Eq.~\eqref{eq:tikhonov_vec_rho} into the definition of $\boldsymbol{c}$, the following linear system is
	found:
	\begin{align}
		\boldsymbol{c} ~ + ~ \frac{1}{\alpha^2} ~ \boldsymbol{M}  ~ \boldsymbol{\Sigma}^{-1} \boldsymbol{c} &= - \boldsymbol{G} \qquad\mbox{ where }\qquad
		\boldsymbol{M} = \boldsymbol{K} \boldsymbol{K}^T \ .
	\end{align}
	Solving this linear system for $\boldsymbol{c}$ at a given value of $\alpha$, the spectral function $\boldsymbol{\rho}$ can be reconstructed using Eq.~(\ref{eq:tikhonov_vec_rho}).
	From the above definitions, it follows that the reconstructed propagator written in terms of $\boldsymbol{c}$ reads
	\begin{equation}
		\boldsymbol{G} =  - ~ \frac{1}{\alpha^2} 	\boldsymbol{M} ~ \boldsymbol{\Sigma}^{-1}  ~ \boldsymbol{c} \ .
		\label{eq:tikhonov_vec_G}
	\end{equation}

	\subsection{The \texorpdfstring{$p^2$}{p2}-formalism}

	Starting from Eq.~\eqref{eq:kl_rep_p2_mu}, the Tikhonov functional \eqref{eq:tikhonov_cov} becomes\footnote{Note that we are not taking into account the correlation between the different momenta, just the variances.}
	\begin{eqnarray}
		J_\alpha & =  & \sum_{i} \frac{1}{\sigma_i^2} \pqty{\int_{\omega_0^2}^\infty
			\frac{\tilde\rho(\mu)}{p_i^2 + \mu} \dd{\mu} - G(p_i)}^2 + \alpha^2 \int_{\omega_0^2}^{\infty} \tilde\rho(\mu)^2 \dd{\mu} \ .
		\label{eq:tikhonov_func_p2}
	\end{eqnarray}
	Upon repeating the functional equivalent of the steps taken above, the following expressions for $\boldsymbol{M}$ and $\boldsymbol{\tilde\rho}$ are found:
	\begin{eqnarray}
		M_{ij} &\coloneqq & \int_{\omega_0^2}^\infty \dd{\mu} \frac{1}{( p_i^2 + \mu) ~ ( p_j^2 + \mu)} =
		\begin{cases}
			\frac{1}{p_j^2 - p_i^2}\ln(\frac{p_j^2 + \omega_0^2}{p_i^2 + \omega_0^2})  & i \neq j \\
			\frac{1}{p_i^2 + \omega_0^2}  & i = j,
		\end{cases}
		\label{eq:M_p2} \\
		& & \nonumber \\
		\tilde\rho(\mu)  & = & - \frac{1}{\alpha^2} \sum_{i} \frac{c_i}{p_i^2 + \mu} \frac{1}{\sigma_i^2} \label{eq:rho_mu} .
	\end{eqnarray}

	\subsection{The \texorpdfstring{$ip$}{ip}-formalism}
	Repeating these steps but starting from Eq.~\eqref{eq:kl_rep_ip} instead, yields
	\begin{equation}
		J_\alpha = \sum_{i} \frac{1}{\sigma_i^2} \pqty{\int_{-\infty}^{\infty} \Omega(\omega, \omega_0) \frac{\rho(\omega)}{\omega - i p_i} \dd{\omega} - G(p_i)}^2 + \alpha^2 \int_{-\infty}^{\infty} \rho(\omega)^2 \dd{\omega},
		\label{eq:tikhonov_func_ip}
	\end{equation}
	resulting in
	\begin{align}
		M_{ij} &\coloneqq \int_{-\infty}^\infty \frac{\dd{\omega}}{\omega - i p_i} \frac{\Omega(\omega, \omega_0)}{\omega - i p_j} =
		\begin{cases}
			2 \,\bqty{\arctan{\frac{p_j}{\omega_0}} - \arctan{\frac{p_i}{\omega_0}} } / ( p_j - p_i ) \, &\mbox{for } i \neq j  \\
			& \\
			2 \, \omega_0 / (p_i^2 + \omega_0^2)  \,  & \mbox{for } i = j,
		\end{cases}
		\label{eq:M_ip} \\
		\rho(\omega) &= - \frac{1}{\alpha^2} \sum_{i} \frac{c_i}{\omega - i p_i} \frac{1}{\sigma_i^2}.
	\end{align}
	It is worth noting that as $\omega_0 \to 0$,
	\begin{align}
		M_{ij}&\to \begin{cases}
			2 \pi/ \left(\abs{p_i} + \abs{p_j} \right)  \, & \mbox{for } p_i \, p_j \leq 0 \ ,\\
			& \\
			0 & \text{otherwise} \ .
			\label{eq:M}
		\end{cases}
	\end{align}
	This is proportional to the $\boldsymbol{M}$ found when inverting a Laplace transform \cite{Laplace}, whereas the spectral representation
	in Eq.~\eqref{eq:kl_rep_p2} can be viewed as a double Laplace transform. This will become relevant when discussing the observed difference in reconstruction quality between the two
	methods. Although both representations are mathematically equivalent, the associated inversion procedures perform differently at the numerical level.

	%========================================================
	%========================================================
	\subsection{Construction of toy models}

	In principle, any of the above methods can be used to compute the spectral function and from it rebuild the propagators. However,
	from the numerical point of view, given the different characteristics of the matrix $\boldsymbol{M} = \boldsymbol{K} \, \boldsymbol{K}^T$, the two
	procedures can behave quite differently.  Therefore, before applying the inversions to the reconstruction of the propagators,
	we investigate their performance on three toy models.

	A Breit-Wigner type model
	\begin{equation}
		\rho(\omega) = \frac{1}{\pi}  \frac{2 \omega \gamma}{(\omega^2 - \gamma^2 - M^2)^2 + 4 \omega^2 \gamma^2},
		\label{eq:rho_bw_german}
	\end{equation}
	with $M=3$ and $\gamma=1$ (dimensionless). This toy model for the spectral function was investigated in \cite{Tripolt:2018xeo}. A similar type of toy model, albeit with a wider peak, was also used in the paper \cite{Dudal:2013yva}.

	A ``Bessel'' model without IR-cutoff,
	\begin{equation}
		\rho(\omega) = \frac{J_1(\omega) J_3(\omega)}{\omega^2},
		\label{eq:rho_bessel}
	\end{equation}
	with $J_n(\omega)$ being the Bessel functions of the first kind, was constructed to obey the same sum rule as gluons and ghosts are supposed to obey, namely
	\begin{equation}\label{eq:sumrule}
		\int_{\omega_0}^{\infty} \rho(\omega) \omega \dd{\omega}=0.
	\end{equation}
	This model is extended to $\omega < 0$ by demanding $\rho(-\omega) = -\rho(\omega)$.
	In Appendix~\ref{A} we have recollected the argument why the spectral functions of the gluons and ghosts, assuming the associated propagators have a Källén-Lehmann spectral representation to begin with, must obey the sum rule \eqref{eq:sumrule}; see \cite{Oehme:1979ai,Oehme:1990kd,Alkofer:2000wg,Cornwall:2013zra} for further reference.
	In short the sum rule \eqref{eq:sumrule} can be obtained using the large momentum behavior of the propagator, whereto, thanks to asymptotic freedom, perturbation theory applies.
	Obviously Eq.~\eqref{eq:sumrule} implies that the spectral function can no longer be positive-definite.
	Consequently, one has to resort to inversion strategies that can accommodate for such spectral functions  \cite{Langfeld:2001cz,Qin:2013ufa,Dudal:2013yva,Rothkopf:2016luz,Ilgenfritz:2017kkp,Tripolt:2018xeo,Cyrol:2018xeq}, which excludes e.g.~the popular (standard) Maximum Entropy Method \cite{Asakawa:2000tr,Aarts:2007pk,Meyer:2011gj}. Alternative methods include Padé rational function approximation \cite{Wang:2018jsp} or machine learning-based methodologies \cite{Fournier}, of which it still needs to be established if these also perform well for unphysical Green functions.
	Yet another recipe for inversion was proposed in \cite{Ferrari:2016snh} based on analytical insights, but to our knowledge it has never been tested in practice, most likely due to the reason mentioned in \cite{Pawlowski:2017gxj}: the required precision to obtain sensible results is unrealistic in numerical computations. Besides these numerical approaches, analytical estimates of spectral functions can also be made. Oftentimes these are performed in conjunction with numerical tools; see \cite{Strauss:2012dg,Cyrol:2018xeq,Boguslavski:2018beu,Meyer:2011gj,Lowdon:2015nha,Lowdon:2017uqe} and references therein for examples.

	A third model, not only obeying Eq.~\eqref{eq:sumrule} but also featuring an IR-cutoff, was constructed with the spectral function
	\begin{align}
		\rho(\omega) &= -\frac{1}{\omega^4 + 4} + \frac{A}{\omega^6 + 2} \text{ for } \omega \geq \sqrt{2} 	\label{eq:rho_david} \ ,
	\end{align}
	where
	\begin{align}
		A &= \frac{3 \pi }{2 \sqrt[3]{2} \left(\pi  \sqrt{3}+ a -2 b +2 \sqrt{3} c \right)} \notag \\
		a &= \log \left(1+2 \sqrt[3]{2}-2^{2/3}\right), \notag \\
		b & = \log \left(1+2^{2/3}\right), \notag \\
		c &= \tan ^{-1}\left(\frac{1- 2^{5/3}}{\sqrt{3}}\right). \notag
	\end{align}
	Again the spectral function was extended to $\omega < 0$ using $\rho(-\omega) = -\rho(\omega)$.

	%========================================================
	%========================================================
	\subsection{Data building and analysis for the toy models}
	\label{sec:data_building}

	In order to mimic the conditions of a lattice simulation, we proceed as follows.
	For a given spectral density (\ref{eq:rho_bw_german}), (\ref{eq:rho_bessel}), (\ref{eq:rho_david}),
	the ``propagator'' $G_\text{orig}$ is computed using either Eq.~\eqref{eq:kl_rep_p2_mu} or Eq.~\eqref{eq:kl_rep_ip}.
	From this $G_\text{orig}$, $N_\text{bootstrap}$ data sets $G_\epsilon$ are generated satisfying a Gaussian distribution
	with mean value $G_\text{orig}$ and variance $(\epsilon G_\text{orig})^2$, i.e. the $G_\epsilon$ are distributed
	according to a probability distribution $G_\epsilon \sim \mathcal{N}(G_\text{orig}, (\epsilon G_\text{orig})^2)$,
	where $\epsilon$ is the noise level (in percentage) of the samples.
	Furthermore, for each data set $\Nres$ momenta are uniformly sampled in the interval $p \in [-10, 10]$ for the $i p$-formalism, of which the range $p \in ~  [0 \, , 10 ]$ are squared for the $p^2$-formalism.
	The $p$ are dimensionless, as there is no physical scale here.
	The choice for this particular range is motivated by the fact that all relevant features of the toy models lie within this range.

	For each of the $G_\epsilon$ bootstrap samples, the inversion is performed using the two formalisms discussed previously. For each $\epsilon$ we have used
	$N_\text{bootstrap}=1000$ samples. In this way the distribution of the optimal $\alpha$ parameter as a function of $\epsilon$ and $\Nres$ could be studied.
	Typically, the distribution of the optimal $\alpha$ is Gaussian or almost Gaussian. The exception occurs for large enough $\Nres$ where the $\alpha$ distributions show a tail
	that touches the point $\alpha = 0$.
	For these small optimal $\alpha$ values, the initial ill-defined inversion reappears, the inversion fails and the original propagator data cannot be reconstructed.
	To prevent these pathological cases we require the optimal $\alpha$ distribution to be compatible with a normal law, i.e.~that they can be fitted by a Gaussian law.
	From the practical point of view, this is enough to prevent the small optimal $\alpha$ values.

	The quality of the reconstructed spectral function $\rho_\text{re}$, defined as the spectral function returned by the inversion methods derived in Eqs. \eqref{eq:tikhonov_func_p2}--\eqref{eq:M}, can be measured from
	the \emph{coefficient of determination} defined as
	\begin{eqnarray}
		R^2 &=& 1 - \frac{\Delta^2_{\text{res}}}{\Delta^2_{\text{tot}}} \label{eq:R2} \ ,
	\end{eqnarray}
	where
	\begin{eqnarray}
		\Delta^2_{\text{res}} &=&  \sum_i \pqty{\rho_\text{orig}(\omega_i) - \rho_\text{re}(\omega_i)}^2, \qquad \Delta_{\text{tot}} = \sum_i \pqty{\rho_\text{orig}(\omega_i) - \bar{\rho}_\text{orig}}^2  \ ,
	\end{eqnarray}
	$\rho_\text{orig}$ are the input data points used to build the propagator and $\bar{\rho}_\text{orig}$ is the mean value of $\rho_\text{orig}$; because
	the function $\rho(\omega)$ is odd, for the evaluation of $R^2$ only data with $\omega \geq 0$ was considered.
	The coefficient of determination measures how the variation of the dependent variable matches the variation of the independent variable.
	It has a maximum value of 1, indicating a perfect fit, but is not bounded from below.

	The quality of the reconstruction could have been measured using a quantity different than $R^2$. For example, in a recent
	work \cite{Tripolt:2018xeo}, the authors defined the applicability of the reconstruction method as the ability to find the position of the dominant peak to within 10\%.
	Therefore, we will provide heat maps for $R^2$ in $(\epsilon, \Nres)$-space onto which solid black contour lines have been drawn to indicate the accuracy in finding the dominant peak position.

	All numerical analysis was performed in the \emph{Python} language, using the \verb|symfit| optimization package \cite{roelfs:2018}.

	\begin{figure*}[t]
		\centering
		\begin{subfigure}[b]{0.49\textwidth}
			\includegraphics[width=\textwidth]{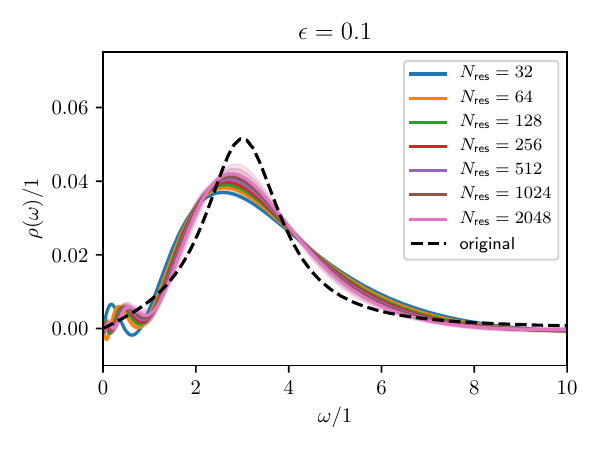}
			\caption{$ip$-method results for $\boldsymbol{\Sigma} \neq \I$}
			\label{fig:bw_fixed_epsilon_rho_ip}
		\end{subfigure}
		\begin{subfigure}[b]{0.49\textwidth}
			\includegraphics[width=\textwidth]{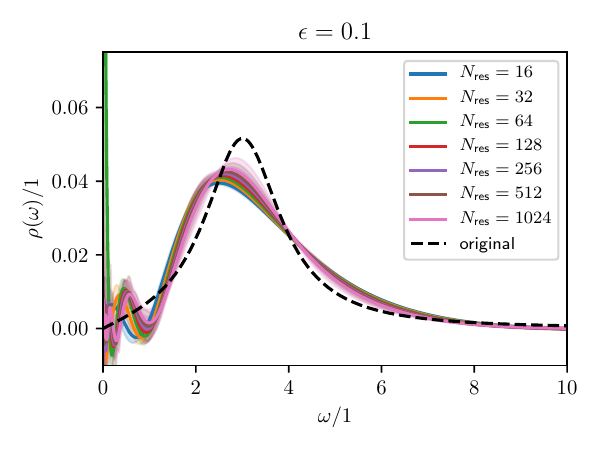}
			\caption{$p^2$-method for $\boldsymbol{\Sigma} \neq \I$}
			\label{fig:bw_fixed_epsilon_rho_p2}
		\end{subfigure}
		\begin{subfigure}[b]{0.49\textwidth}
			\includegraphics[width=\textwidth]{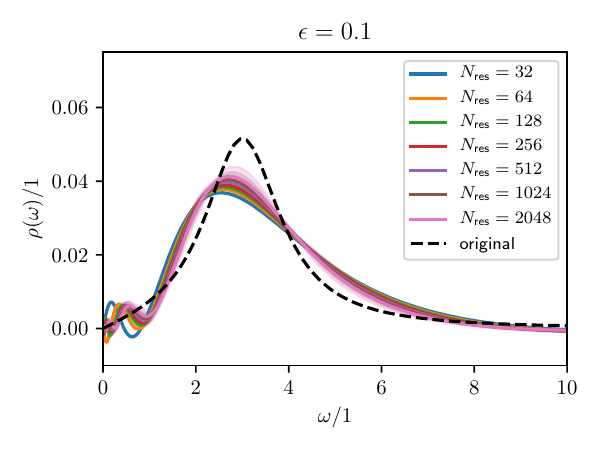}
			\caption{$ip$-method results with $\boldsymbol{\Sigma} = \I$}
			\label{fig:bw_fixed_epsilon_rho_ip_unweighted}
		\end{subfigure}
		\begin{subfigure}[b]{0.49\textwidth}
			\includegraphics[width=\textwidth]{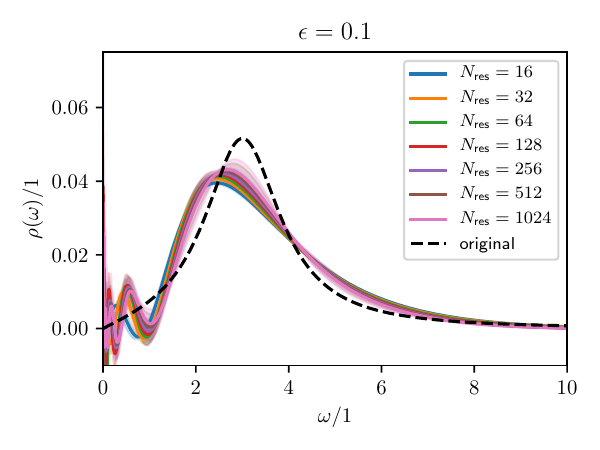}
			\caption{$p^2$-method results with $\boldsymbol{\Sigma} = \I$}
			\label{fig:bw_fixed_epsilon_rho_p2_unweighted}
		\end{subfigure}
		\caption{Spectral function from the inversion as a function of $\Nres$ at a fixed noise of $\epsilon=0.1\%$ for the Breit-Wigner model. Here, $\omega$ is a dimensionless quantity. The inversion using the $p^2$-method returns a spectral density with larger oscillations when $\omega \to 0$. More on the origin of this in Section \ref{sec:p2_vs_ip}}.
		\label{fig:bw_fixed_epsilon}
	\end{figure*}

	\subsection{Determination of \texorpdfstring{$\omega_0$}{omega}}\label{sec:determination_of_cutoff}

	In general, a physical cutoff $\omega_0$ should ideally not depend on the choice of regularization, in particular it should not depend on our choice of the Tikhonov parameter $\alpha$. This means that the variation of $\omega_0$ w.r.t.~$\alpha$ should be as small as possible in practice. Since numerically we have easier access to the variation of $\alpha$ w.r.t.~$\omega_0$, the optimum value for $\alpha$ is more easily identified as the regions where the variation of $\alpha$ w.r.t.~$\omega_0$ is \emph{maximal}.

	This means that after generating the $\omega_0$ v.s.~$\alpha$ curve from our bootstrap, the point with the largest standard deviation in $\alpha$ w.r.t. $\omega_0$ corresponds to the most likely physical cutoff $\omega_0$.

	This concept will guide our choices of the optimal $\omega_0$ and hence, the corresponding $\alpha(\omega_0)$.

	%=================================================================
	%=================================================================
	\section{Toy model---results and discussion}
	\label{sec:toy_model_results}

	For all of the toy models described in the previous section, the $p^2$- and $ip$-formalism derived in Equations \eqref{eq:tikhonov_func_p2}--\eqref{eq:M_ip} will be
	applied to data sets generated as described in Section \ref{sec:data_building}, while setting $\Sigma_{ij}$ to either $\sigma_i^2 \delta_{ij}$ or $\delta_{ij}$ during the inversion, where $\sigma_i^2$ is the variance of $G(p_i)$.

	%=================================================================
	%=================================================================
	\subsection{The Breit-Wigner Spectral Function}
	\label{sec:BWresults}

	We start our analysis by looking at the inverse problem for the Breit-Wigner type model given in Eq.~\eqref{eq:rho_bw_german}.
	This model has no IR-cutoff and the inversions performed here therefore consider $\omega_0 = 0$. This implies that the $p=0$ data point has to be excluded from the
	inversion for both methods. The effect of a non-zero $\omega_0$ will be studied later on.

	In Figure~\ref{fig:bw_fixed_epsilon}, the effect of $\Nres$ at fixed noise $\epsilon = 0.1 \%$ is shown for the various methods.
	A comparison of Figure~\ref{fig:bw_fixed_epsilon_rho_ip} with Figure \ref{fig:bw_fixed_epsilon_rho_ip_unweighted}, and Figure \ref{fig:bw_fixed_epsilon_rho_p2} with Figure
	\ref{fig:bw_fixed_epsilon_rho_p2_unweighted}, shows that the effects of taking into account the error on the data, i.e.~setting $\boldsymbol{\Sigma} \neq \I$,
	does not visibly change the quality of the spectral function computed from the inversion.
	Furthermore, Figures~\ref{fig:bw_fixed_epsilon_rho_ip} and \ref{fig:bw_fixed_epsilon_rho_p2} show that the $ip$-method performs better for this toy model at low momentum scales, as
	the spectral function evaluated with the $p^2$-method has larger oscillations at small $\omega$. For the $p^2$-method, the IR oscillations are reduced by increasing $\Nres$.
	The different IR behaviour of the two methods will be discussed quantitatively below.

	Another important feature that can be observed from Figure~\ref{fig:bw_fixed_epsilon} is that for the level of noise considered, $\epsilon = 0.1\%$, the inversions do
	not have a strong dependence on the number of data points, $\Nres$, taken into account in the inversion.
	Although the computed spectral functions coming from the inversions are not perfect, it seems that both methods capture the main features of $\rho ( \omega )$ for any $\Nres$.
	As discussed below, it is the value of $\epsilon$ that seems to play the most important role in the inversion, with the inverted spectral function getting closer
	to the exact spectral function as $\epsilon$ is reduced, as expected. Note also that despite the dependence on $\Nres$ of the computation of $\rho$ is mild, the match between the computed and input spectral function slightly improves as $\Nres$ increases.

	\begin{figure*}[t]
		\centering
		\begin{subfigure}[b]{0.49\textwidth}
			\includegraphics[width=\textwidth]{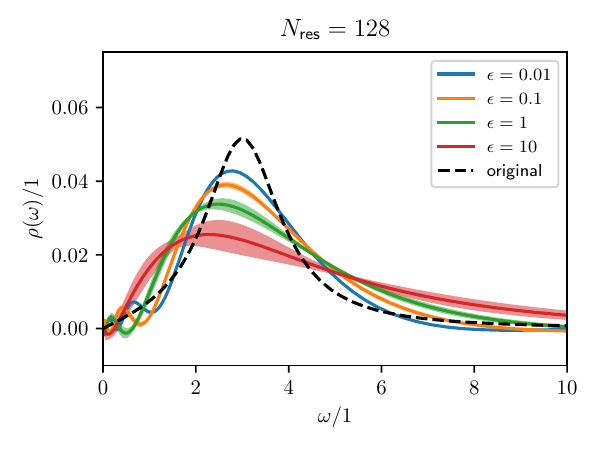}
			\caption{$ip$-method results for $\boldsymbol{\Sigma} \neq \I$}
			\label{fig:bw_fixed_Nres_rho_ip}
		\end{subfigure}
		\begin{subfigure}[b]{0.49\textwidth}
			\includegraphics[width=\textwidth]{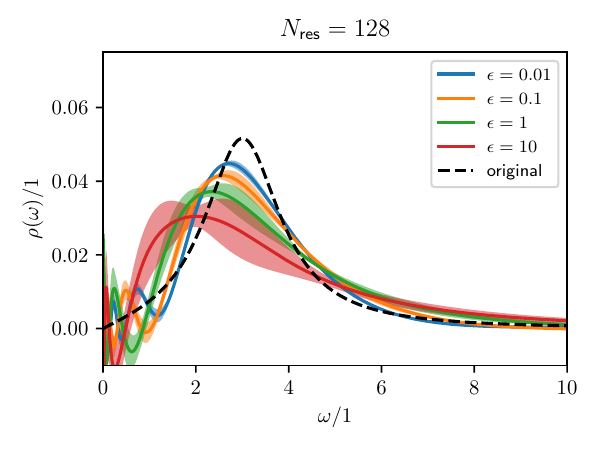}
			\caption{$p^2$-method results for $\boldsymbol{\Sigma} \neq \I$}
			\label{fig:bw_fixed_Nres_rho_p2}
		\end{subfigure}
		\begin{subfigure}[b]{0.49\textwidth}
			\includegraphics[width=\textwidth]{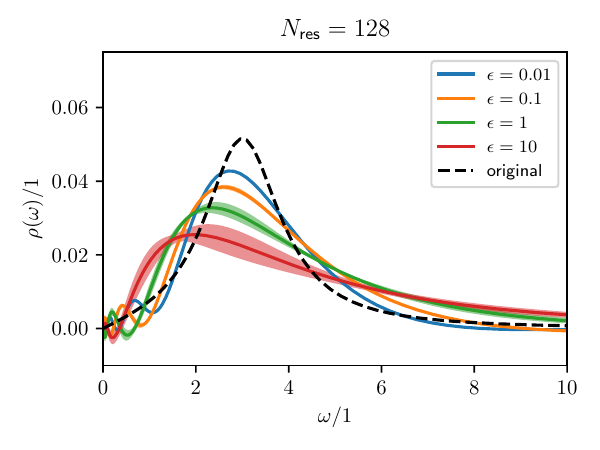}
			\caption{$ip$-method results with $\boldsymbol{\Sigma} = \I$}
			\label{fig:bw_fixed_Nres_rho_ip_unweighted}
		\end{subfigure}
		\begin{subfigure}[b]{0.49\textwidth}
			\includegraphics[width=\textwidth]{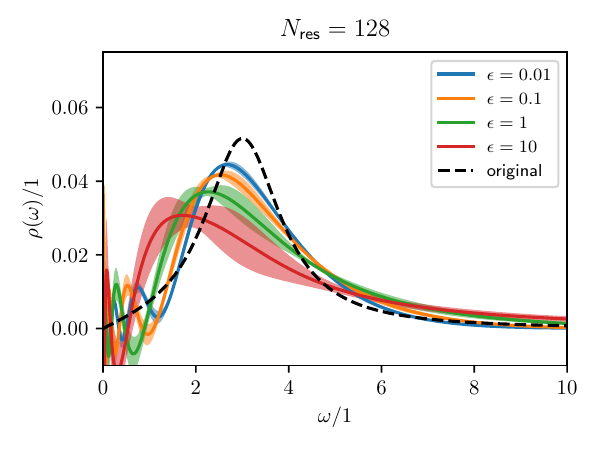}
			\caption{$p^2$-method results with $\boldsymbol{\Sigma} = \I$}
			\label{fig:bw_fixed_Nres_rho_p2_unweighted}
		\end{subfigure}
		\caption{Spectral function from the inversion as a function of $\epsilon$ for $\Nres=128$ for the Breit-Wigner model. Here, $\omega$ is a dimensionless quantity.}
		\label{fig:bw_fixed_Nres}
	\end{figure*}

	The effects of the noise level $\epsilon$ is shown in Figure~\ref{fig:bw_fixed_Nres} for the two methods and for $\Nres=128$. Other values of $\Nres$
	show similar results. Again, taking into account the statistical errors results in a
	reconstructed $\rho$ with smaller errors. In general, reducing the noise level results in a computed spectral function
	that is closer to the exact $\rho ( \omega )$; see Figure~\ref{fig:bw_fixed_Nres_rho_ip} against \ref{fig:bw_fixed_Nres_rho_ip_unweighted}, and compare
	Figure~\ref{fig:bw_fixed_Nres_rho_p2} with \ref{fig:bw_fixed_Nres_rho_p2_unweighted}.
	Additionally, as shown in Figures \ref{fig:bw_fixed_Nres_rho_ip} and \ref{fig:bw_fixed_Nres_rho_p2}, the $ip$-method provides a spectral function that is less oscillatory in the
	IR. This is also evidenced by the corresponding increase in the standard deviation for $\rho$,
	suggesting that the $p^2$-reconstructions are less IR stable. However, concerning the location and height of the maximum of the computed
	$\rho ( \omega )$, the $i p$-method captures the location best, while the $p^2$-method seems to better captures the height of the maximum.
	\begin{figure*}[t]
		\centering
		\begin{subfigure}[b]{0.49\textwidth}
			\includegraphics[width=\textwidth]{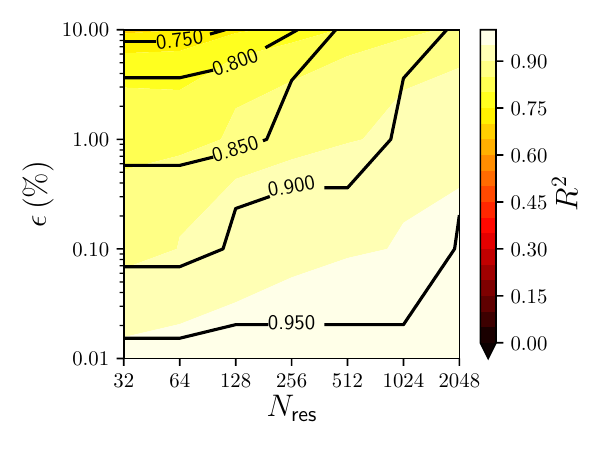}
			\caption{$ip$-method results for $\boldsymbol{\Sigma} \neq \I$}
			\label{fig:bw_R2_rho_ip}
		\end{subfigure}
		\begin{subfigure}[b]{0.49\textwidth}
			\includegraphics[width=\textwidth]{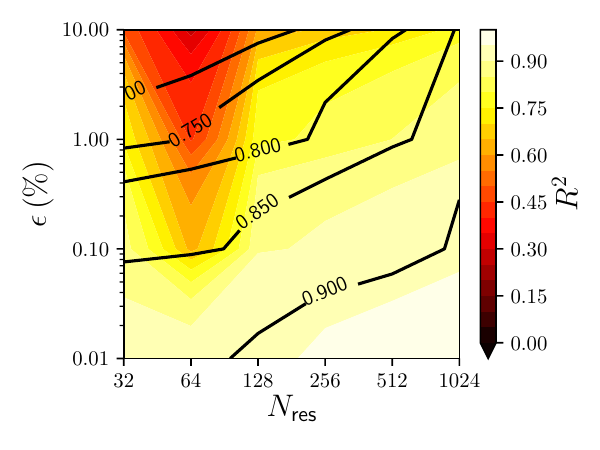}
			\caption{$p^2$-method results for $\boldsymbol{\Sigma} \neq \I$}
			\label{fig:bw_R2_rho_p2}
		\end{subfigure}
		\begin{subfigure}[b]{0.49\textwidth}
			\includegraphics[width=\textwidth]{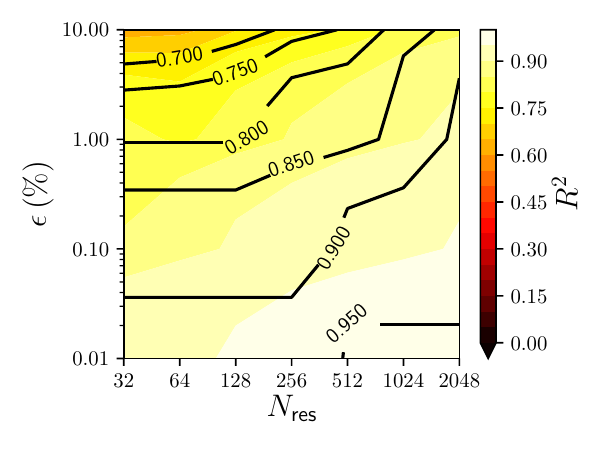}
			\caption{$ip$-method results for $\boldsymbol{\Sigma} = \I$}
			\label{fig:bw_R2_rho_ip_unweighted}
		\end{subfigure}
		\begin{subfigure}[b]{0.49\textwidth}
			\includegraphics[width=\textwidth]{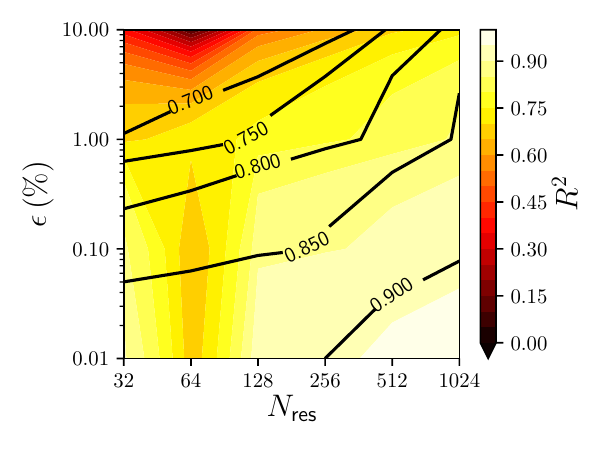}
			\caption{$p^2$-method results for $\boldsymbol{\Sigma} = \I$}
			\label{fig:bw_R2_rho_p2_unweighted}
		\end{subfigure}
		\caption{$R^2$ as a function of $\epsilon$ and $\Nres$ for the Breit-Wigner model. Solid black contour lines indicate the accuracy in finding the dominant peak position.}
		\label{fig:bw_R2}
	\end{figure*}

	The behaviour of the coefficient of determination $R^2$ as a function of the noise level $\epsilon$ and number of momenta $\Nres$ is summarized in
	Figure~\ref{fig:bw_R2}. On the figure the solid black contour lines indicate the accuracy in finding the dominant peak position, i.e.~the relative error on the momentum position of the maximum of the spectral function.  The results in this figure indicate that both methods perform better when the statistical errors on the propagator data are taken into account in the inversion. In general, the value of $R^2$ is closer to unity for the $i p$-method. In particular for the smaller values of $\Nres$,
	the $p^2$-method oftentimes  has $R^2 \leqslant 0.5$. These results for $R^2$ can be viewed as an indication that overall the solution provided by the $i p$-method is closer to the original spectral function.

	As can be seen in Figure~\ref{fig:bw_R2} it is the value of $\epsilon$ that has a major impact on the reconstruction of the spectral function.
	If, for example, $R^2 > 0.9$ is demanded, a noise level of about $\epsilon \lesssim 0.1\%$ is needed for the $ip$-method to fulfil this condition,
	while the $p^2$-method requires an $\epsilon \lesssim 0.05\%$ to achieve the same values of the coefficient of determination $R^2$.

	The spectral function under analysis was also investigated in \cite{Tripolt:2018xeo} using various inversion techniques namely Maximum Entropy (MEM), Backus-Gilbert (BG)
	and the Schlessinger point method (SP).
	According to the authors of \cite{Tripolt:2018xeo}, the MEM, BG and SP are able to locate the maximum of $\rho$ quite well, both its position and height.
	The SP provides the best reconstructed spectral function but not necessarily for the smallest errors (see their Figure~6). A fair comparison is difficult to perform.
	Their spectral functions that were reconstructed using BG are too broad and clearly quite far away from the input $\rho ( \omega )$ (see their Figure~5).
	Their implementation of the MEM returns
	a spectral function with large oscillations at small momentum scales, i.e.~for $\omega \lesssim 220$ MeV, and although it provides a good description of the position
	of the maximum of $\rho$, it clearly underestimates its strength. Even though \cite{Tripolt:2018xeo} does not calculate $R^2$, their graphs suggest that the corresponding $R^2$ values would be smaller due to lesser overlap between the original and the reconstructions as $\Nres$ decreases or $\epsilon$ increases.
	The best choice of algorithm therefore seems to depend on the desired feature of the data we are trying to capture.

	Let us now discuss the effects due to a non-vanishing IR-cutoff $\omega_0$ on the inversion.
	Recall that, as discussed in Section \ref{sec:determination_of_cutoff}, the physical cutoff $\omega_0$ should correspond to the point where the variation of $\alpha$ w.r.t.~$\omega_0$ is maximal.

\begin{figure*}[htb]
		\centering
		\begin{subfigure}[b]{0.49\textwidth}
			\includegraphics[width=\textwidth]{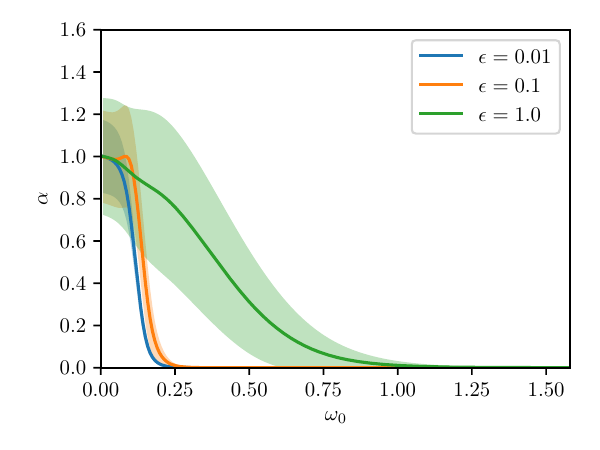}
			\caption{$ip$-method, including $\boldsymbol{\Sigma}$}
			\label{fig:bw_omega0_vs_alpha_ip}
		\end{subfigure}
		\begin{subfigure}[b]{0.49\textwidth}
			\includegraphics[width=\textwidth]{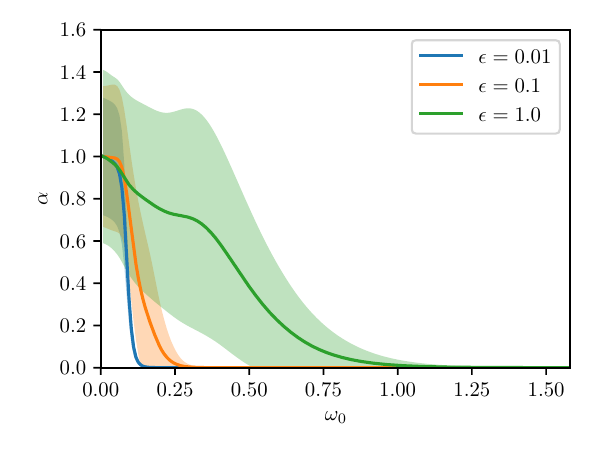}
			\caption{$p^2$-method, including $\boldsymbol{\Sigma}$}
			\label{fig:bw_omega0_vs_alpha_p2}
		\end{subfigure}
		\begin{subfigure}[b]{0.49\textwidth}
			\includegraphics[width=\textwidth]{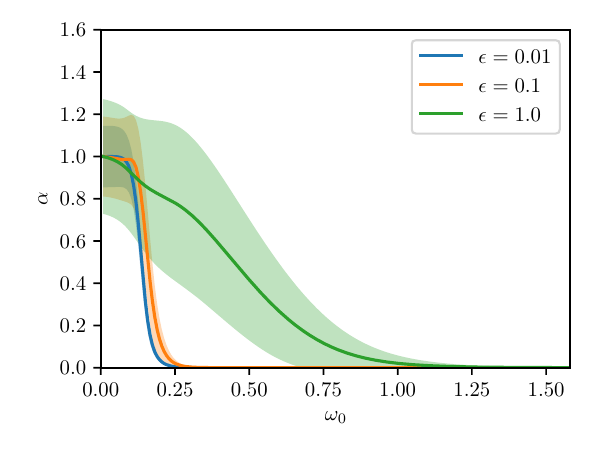}
			\caption{$ip$-method, $\boldsymbol{\Sigma} = \I$}
			\label{fig:bw_omega0_vs_alpha_ip_unweighted}
		\end{subfigure}
		\begin{subfigure}[b]{0.49\textwidth}
			\includegraphics[width=\textwidth]{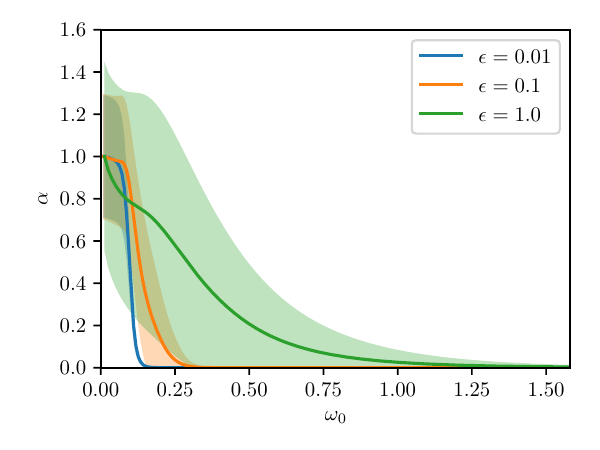}
			\caption{$p^2$-method, $\boldsymbol{\Sigma} = \I$}
			\label{fig:bw_omega0_vs_alpha_p2_unweighted}
		\end{subfigure}
		\caption{$\omega_0$ vs.~$\alpha$ for the Breit-Wigner model at $\Nres=128$.}
		\label{fig:bw_omega0_vs_alpha}
	\end{figure*}

	As a first example, the current Breit-Wigner model has no cutoff. As can be observed in Figure~\ref{fig:bw_omega0_vs_alpha}
	for $\Nres = 128$, in the low noise limit with both methods the maximal standard deviation is achieved around the maximum of the $\omega_0$ v.s.~$\alpha$ curve, suggesting that $\omega_0 = 0$ is indeed the right choice.

	\clearpage
	\newpage

	%===================================================================================
	%===================================================================================
	\subsection{The Bessel spectral function}

	In this section we discuss the results for the inversion when the input function used to generate the propagator data is the
	Bessel model given in Eq.~\eqref{eq:rho_bessel}. The interest in this type of function arises from the property
	\begin{displaymath}
		\int_0^{\infty} \rho(\omega) \omega \dd{\omega} = 0 \ ,
	\end{displaymath}
	i.e. $\rho ( \omega )$ necessarily has regions where it takes positive and negative values and, therefore, mimics a spectral function that
	can be associated with unphysical particles. Firstly, we consider the inversion with $\omega_0 = 0$, which requires
	excluding the $p=0$ data point to avoid the singularity in the matrix $\boldsymbol{M}$. Later on we also look at the case where
	a finite IR cutoff is present.

	\begin{figure*}[bt]
		\centering
		\begin{subfigure}[b]{0.49\textwidth}
			\includegraphics[width=\textwidth]{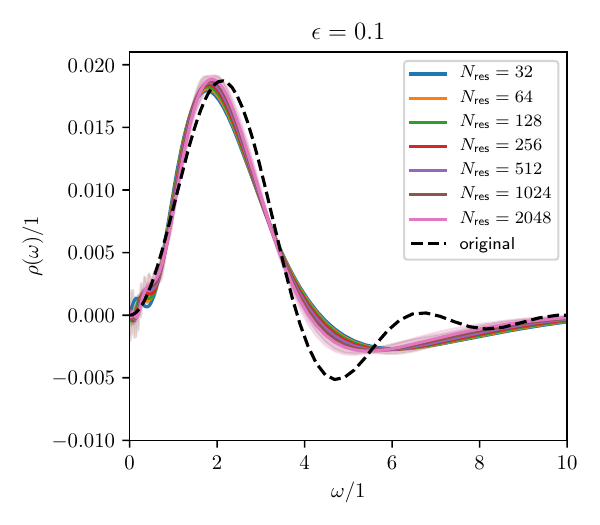}
			\caption{$ip$-method with $\boldsymbol{\Sigma} \neq \I$}
			\label{fig:bessel_fixed_epsilon_rho_ip}
		\end{subfigure}
		\begin{subfigure}[b]{0.49\textwidth}
			\includegraphics[width=\textwidth]{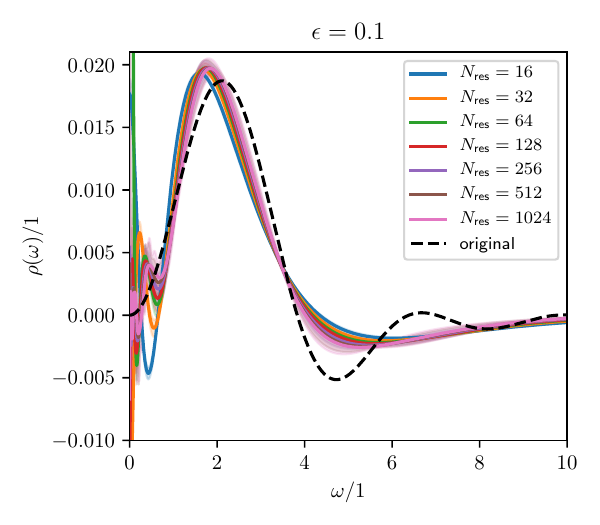}
			\caption{$p^2$-method with $\boldsymbol{\Sigma} \neq \I$}
			\label{fig:bessel_fixed_epsilon_rho_p2}
		\end{subfigure}
		\begin{subfigure}[b]{0.49\textwidth}
			\includegraphics[width=\textwidth]{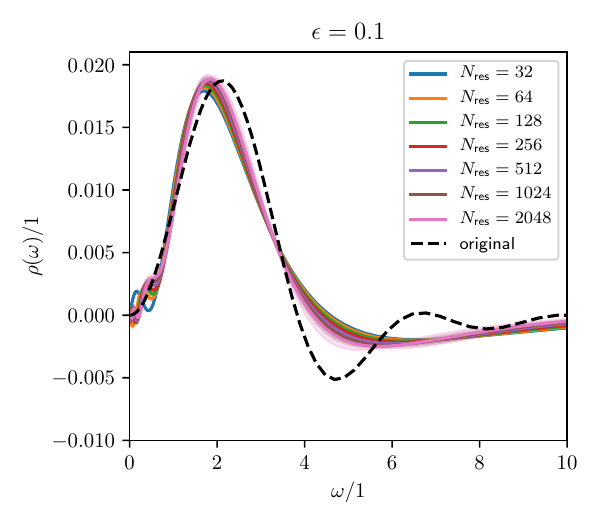}
			\caption{$ip$-method with $\boldsymbol{\Sigma} = \I$}
			\label{fig:bessel_fixed_epsilon_rho_ip_unweighted}
		\end{subfigure}
		\begin{subfigure}[b]{0.49\textwidth}
			\includegraphics[width=\textwidth]{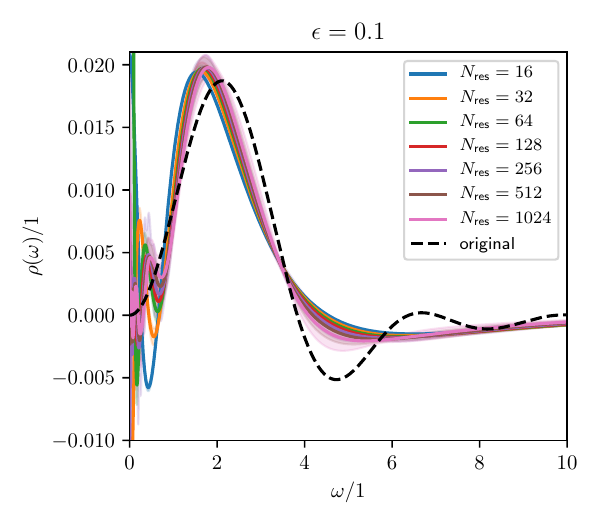}
			\caption{$p^2$-method with $\boldsymbol{\Sigma} = \I$}
			\label{fig:bessel_fixed_epsilon_rho_p2_unweighted}
		\end{subfigure}
		\caption{The reconstructed $\rho ( \omega )$ for various $\Nres$ and for a fixed noise level $\epsilon=0.1\%$ for the Bessel spectral function.}
		\label{fig:bessel_fixed_epsilon}
	\end{figure*}

	\begin{figure*}[tb]
		\centering
		\begin{subfigure}[b]{0.49\textwidth}
			\includegraphics[width=\textwidth]{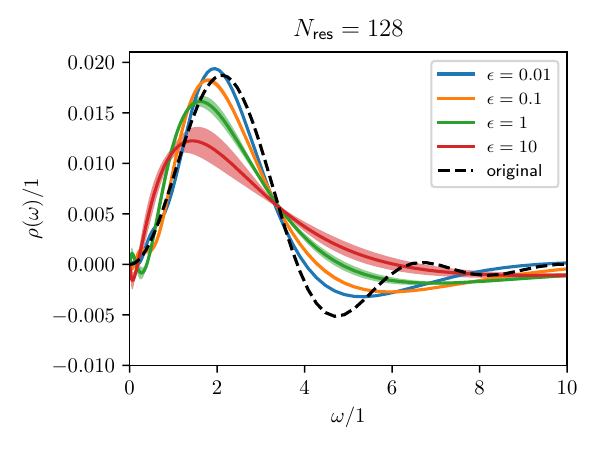}
			\caption{$ip$-method with $\boldsymbol{\Sigma} \ne \I$}
			\label{fig:bessel_fixed_Nres_rho_ip}
		\end{subfigure}
		\begin{subfigure}[b]{0.49\textwidth}
			\includegraphics[width=\textwidth]{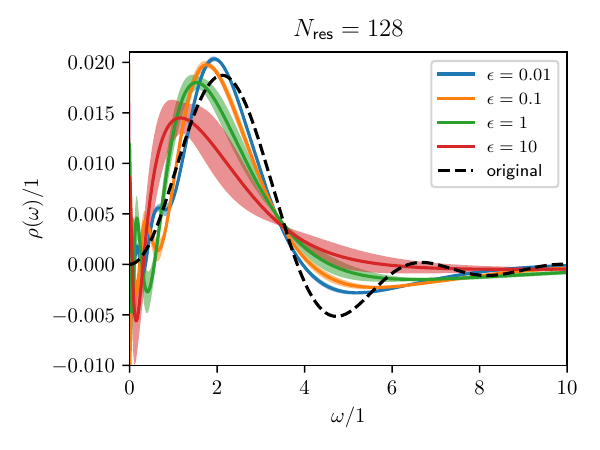}
			\caption{$p^2$-method with $\boldsymbol{\Sigma} \ne \I$}
			\label{fig:bessel_fixed_Nres_rho_p2}
		\end{subfigure}
		\begin{subfigure}[b]{0.49\textwidth}
			\includegraphics[width=\textwidth]{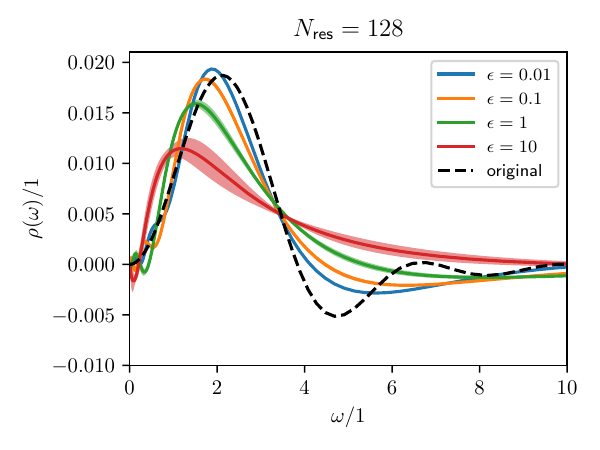}
			\caption{$ip$-method with $\boldsymbol{\Sigma} = \I$}
			\label{fig:bessel_fixed_Nres_rho_ip_unweighted}
		\end{subfigure}
		\begin{subfigure}[b]{0.49\textwidth}
			\includegraphics[width=\textwidth]{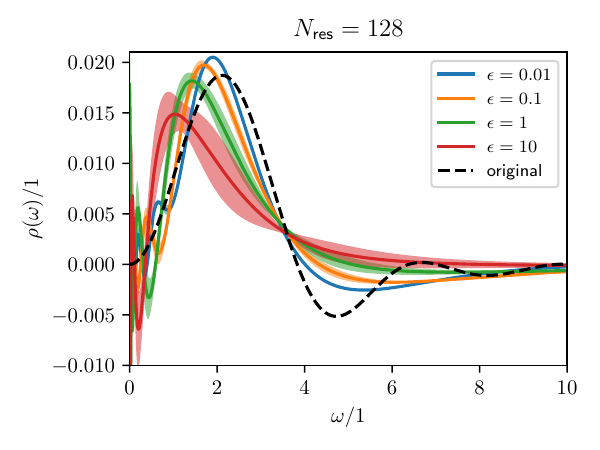}
			\caption{$p^2$-method with $\boldsymbol{\Sigma} = \I$}
			\label{fig:bessel_fixed_Nres_rho_p2_unweighted}
		\end{subfigure}
		\caption{The reconstructed $\rho ( \omega ) $ as a function of $\epsilon$ at a fixed number of momenta $\Nres=128$ for the Bessel model.}
		\label{fig:bessel_fixed_Nres}
	\end{figure*}

		In Figure \ref{fig:bessel_fixed_epsilon}, the effect of $\Nres$ at fixed noise of $\epsilon = 0.1 \%$ is shown for the various methods.
	Similar to the Breit-Wigner model, it seems that including the $\boldsymbol{\Sigma}$-matrix has no visible advantage at this noise level; compare Figure~\ref{fig:bessel_fixed_epsilon_rho_ip} with \ref{fig:bessel_fixed_epsilon_rho_ip_unweighted}, and Figure~\ref{fig:bessel_fixed_epsilon_rho_p2} with
	\ref{fig:bessel_fixed_epsilon_rho_p2_unweighted}.
	The comparison of Figure~\ref{fig:bessel_fixed_epsilon_rho_ip} with Figure~\ref{fig:bessel_fixed_epsilon_rho_p2} shows, again, much greater oscillatory behaviour in
	the IR associated with the $p^2$-method. As for the Breit-Wigner model, the inversion shows only a mild dependence on $\Nres$, with the main
	effect associated with an increase in the number of momenta included in the inversion being a reduction of the statistical errors.
	Both methods are able to locate the maximum of $\rho ( \omega )$ rather well, with the $ip$-method performing slightly better.  Moreover, both methods struggle to reproduce the oscillatory tail of the model for $\omega \gtrsim 4$.

	In Figure~\ref{fig:bessel_fixed_Nres}, the effect of the noise level on the inversion is shown for $\Nres = 128$.
	Again, as for the Breit-Wigner model discussed in Section~\ref{sec:BWresults}, the reconstructed spectral function becomes closer to the input
	$\rho ( \omega )$ when $\epsilon$ is reduced.
	By comparing Figures~\ref{fig:bessel_fixed_Nres_rho_ip} and \ref{fig:bessel_fixed_Nres_rho_ip_unweighted}, and Figures~\ref{fig:bessel_fixed_Nres_rho_p2} and \ref{fig:bessel_fixed_Nres_rho_p2_unweighted}, it can be observed that including the $\boldsymbol{\Sigma}$-matrix gives a minor advantage in the IR as the noise level increases.
	Additionally, as Figures~\ref{fig:bessel_fixed_Nres_rho_ip} and \ref{fig:bessel_fixed_Nres_rho_p2} show, the IR behaviour of the $ip$-method is less oscillatory than that of the
	$p^2$-method.

	\begin{figure*}[h]
		\centering
		\begin{subfigure}[b]{0.49\textwidth}
			\includegraphics[width=\textwidth]{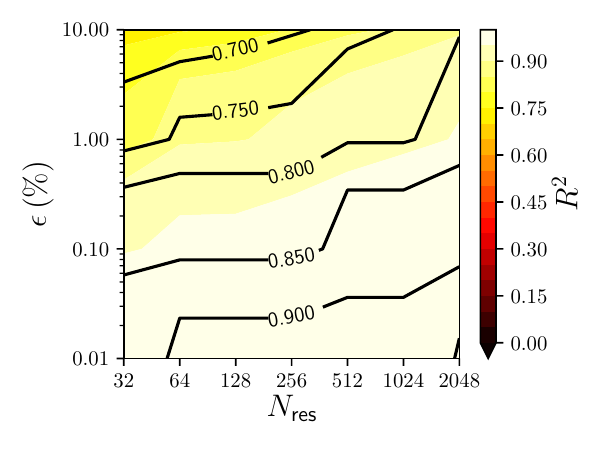}
			\caption{$ip$-method with $\boldsymbol{\Sigma} \ne \I$}
			\label{fig:bessel_R2_rho_ip}
		\end{subfigure}
		\begin{subfigure}[b]{0.49\textwidth}
			\includegraphics[width=\textwidth]{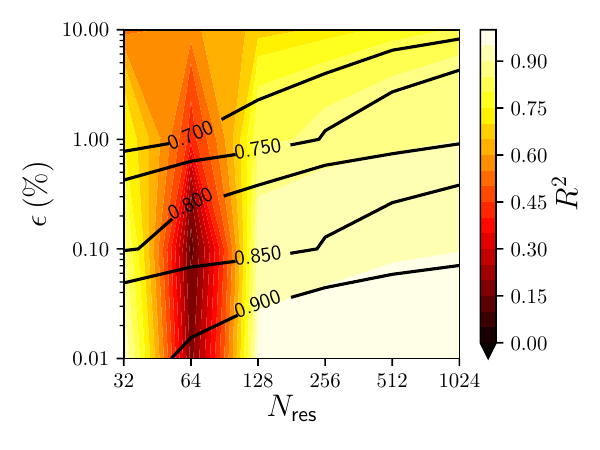}
			\caption{$p^2$-method with $\boldsymbol{\Sigma} \ne \I$}
			\label{fig:bessel_R2_rho_p2}
		\end{subfigure}
		\begin{subfigure}[b]{0.49\textwidth}
			\includegraphics[width=\textwidth]{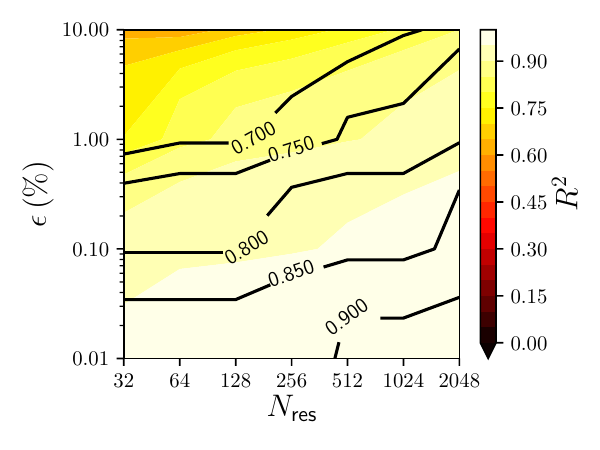}
			\caption{$ip$-method with $\boldsymbol{\Sigma} = \I$}
			\label{fig:bessel_R2_rho_ip_unweighted}
		\end{subfigure}
		\begin{subfigure}[b]{0.49\textwidth}
			\includegraphics[width=\textwidth]{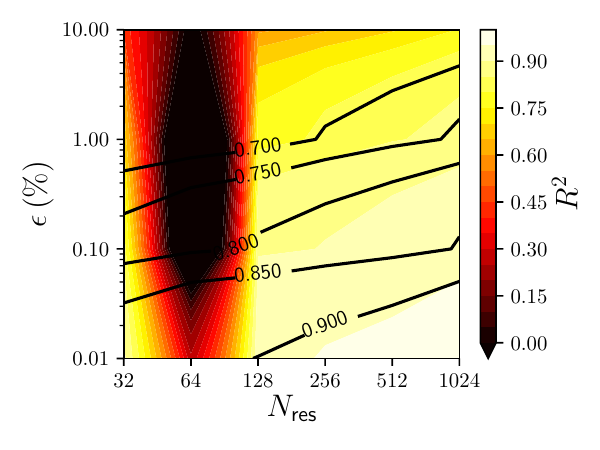}
			\caption{$p^2$-method with $\boldsymbol{\Sigma} = \I$}
			\label{fig:bessel_R2_rho_p2_unweighted}
		\end{subfigure}
		\caption{$R^2$ in the $(\epsilon, \Nres)$-space for the Bessel model. Solid black contour lines indicate the relative error  in finding the dominant peak position.}
		\label{fig:bessel_R2}
	\end{figure*}

	Figure \ref{fig:bessel_R2} shows how $R^2$, as defined in Eq.~\eqref{eq:R2}, changes as the resolution $\Nres$ and the noise $\epsilon$ are varied.
	As can be observed, both methods perform better when the $\boldsymbol{\Sigma}$-matrix is included.
	In general, in order to reproduce a given $R^2$, the $i p$-method allows for a larger noise value than the $p^2$-method.
	For example, if one requires an $R^2 > 0.9$, a noise level of about $\epsilon \lesssim 0.1\%$ is needed for the $ip$-method, while
	$\epsilon \lesssim 0.05\%$ is required for the $p^2$-method.
	The dark areas in Figure \ref{fig:bessel_R2} appearing for the $p^2$-method with $\Nres = 64$ and where $R^2 \approx 0$
	are due to the large IR mismatch. Indeed, as can be seen in Figure~\ref{fig:bessel_fixed_epsilon}, for this particular inversion the $p^2$-method
	returns a spectral function highly oscillatory in the IR that is far from the original function.
	Looking at the effect of the parameters $\epsilon$ and $\Nres$, it is clear that $\epsilon$ has a much larger effect on the quality of reconstruction than $\Nres$.

\begin{figure*}[tbh]
		\centering
		\begin{subfigure}[b]{0.49\textwidth}
			\includegraphics[width=\textwidth]{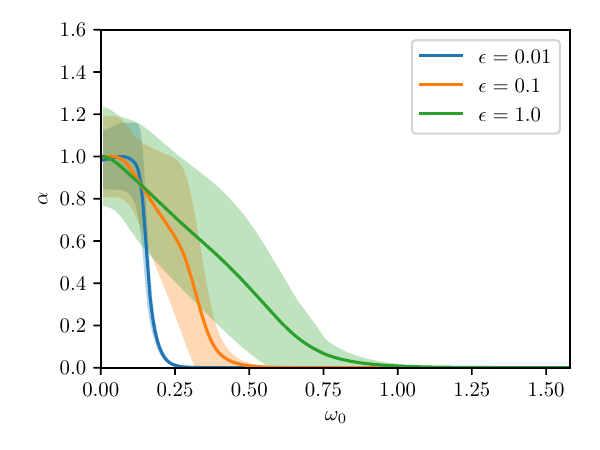}
			\caption{$ip$-method with $\boldsymbol{\Sigma} \ne \I$}
			\label{fig:bessel_omega0_vs_alpha_ip}
		\end{subfigure}
		\begin{subfigure}[b]{0.49\textwidth}
			\includegraphics[width=\textwidth]{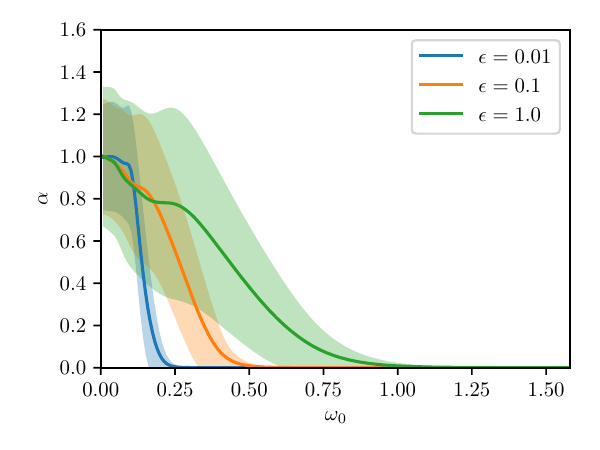}
			\caption{$p^2$-method with $\boldsymbol{\Sigma} \ne \I$}
			\label{fig:bessel_omega0_vs_alpha_p2}
		\end{subfigure}
		\begin{subfigure}[b]{0.49\textwidth}
			\includegraphics[width=\textwidth]{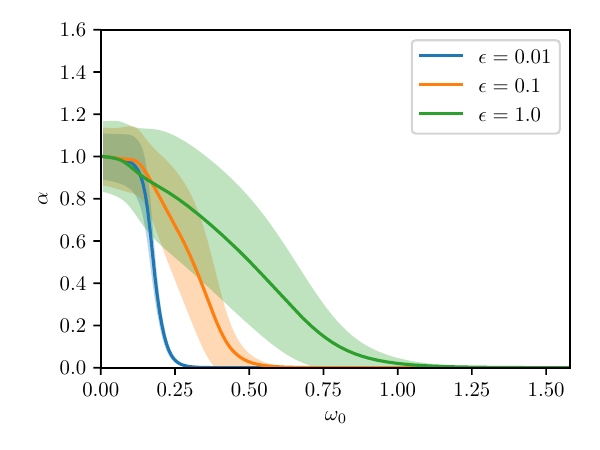}
			\caption{$ip$-method with $\boldsymbol{\Sigma} = \I$}
			\label{fig:bessel_omega0_vs_alpha_ip_unweighted}
		\end{subfigure}
		\begin{subfigure}[b]{0.49\textwidth}
			\includegraphics[width=\textwidth]{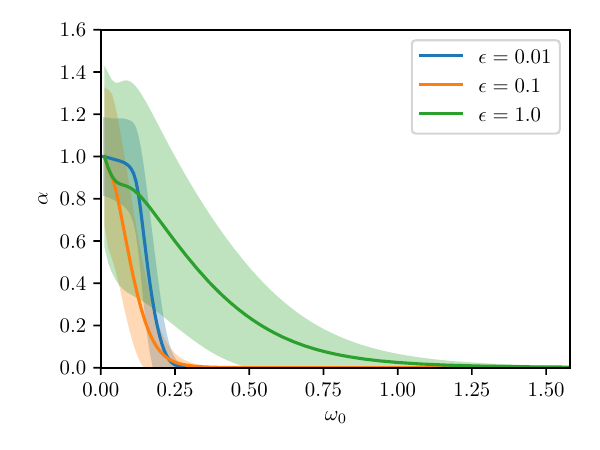}
			\caption{$p^2$-method with $\boldsymbol{\Sigma} = \I$}
			\label{fig:bessel_omega0_vs_alpha_p2_unweighted}
		\end{subfigure}
		\caption{$\omega_0$ vs.~$\alpha$ for the Bessel model at $\Nres=128$.}
		\label{fig:bessel_omega0_vs_alpha}
	\end{figure*}

	As for the Breit-Wigner spectral function, we also investigated how the inversion performs when an IR-cutoff $\omega_0$ is introduced. In Figure~\ref{fig:bessel_omega0_vs_alpha}  we report the $\omega_0$ versus $\alpha$ curve for $\Nres = 128$ at various noise levels. A pattern analogous to that of the Breit-Wigner case emerges, with $\omega_0 = 0$ being suggested as a good candidate.

	The spectral function (\ref{eq:rho_bessel}) was designed to obey the sum rule \eqref{eq:sumrule}.
	However, in practice the numerical integral over a large momentum range tends to diverge as the reconstruction at higher momenta does not go to zero faster
	than $1/p^2$ as expected, but stays small and finite. We found that this can remedied by imposing the sum rule as a constraint on the minimizing functional.
	The constraint does not change the IR, but forces the UV tail to zero as $p^2 \to \infty$ in such a way that the sum rule is satisfied. Such a constraint fit is numerically
	more expensive and, therefore, we choose here to focus on the bootstrap results. The analysis of the constrained fit will be the subject of a future publication.

	\subsection{The spectral function for a model with a cutoff}
	\label{Sec:SFwithcut}
	We now consider the toy model (\ref{eq:rho_david}) which has a physical IR cutoff $\omega_0^*=\sqrt 2$. We start the discussion by looking at the dependence of the inversion on $\omega_0$. Ideally, the $\omega_0$ we determine during the inversion will coincide with $\omega_0^*$.

\begin{figure*}[tbh]
		\centering
		\begin{subfigure}[b]{0.49\textwidth}
			\includegraphics[width=\textwidth]{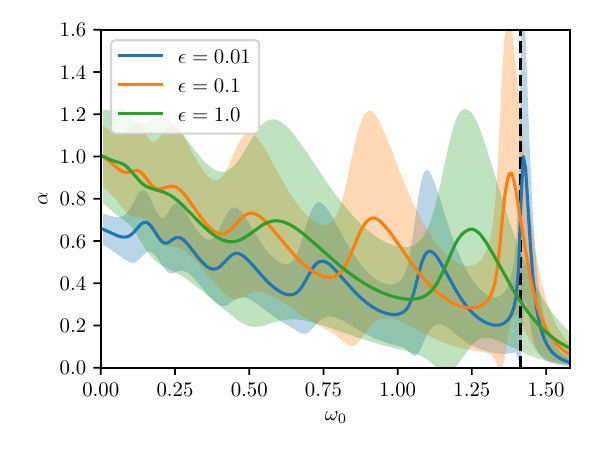}
			\caption{$ip$-method with $\boldsymbol{\Sigma} \ne \I$}
			\label{fig:cutoff_omega0_vs_alpha_ip}
		\end{subfigure}
		\begin{subfigure}[b]{0.49\textwidth}
			\includegraphics[width=\textwidth]{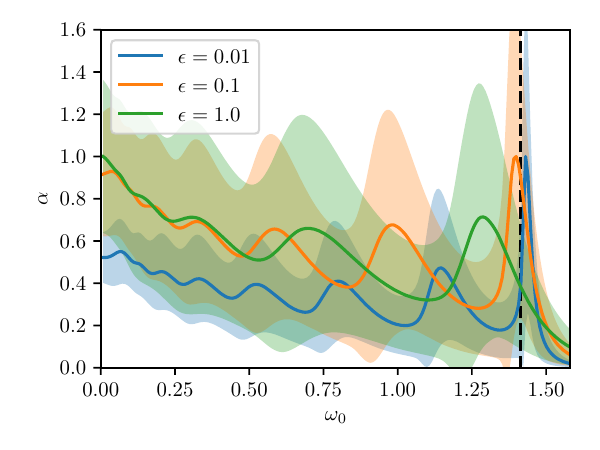}
			\caption{$p^2$-method with $\boldsymbol{\Sigma} \ne \I$}
			\label{fig:cutoff_omega0_vs_alpha_p2}
		\end{subfigure}
		\begin{subfigure}[b]{0.49\textwidth}
			\includegraphics[width=\textwidth]{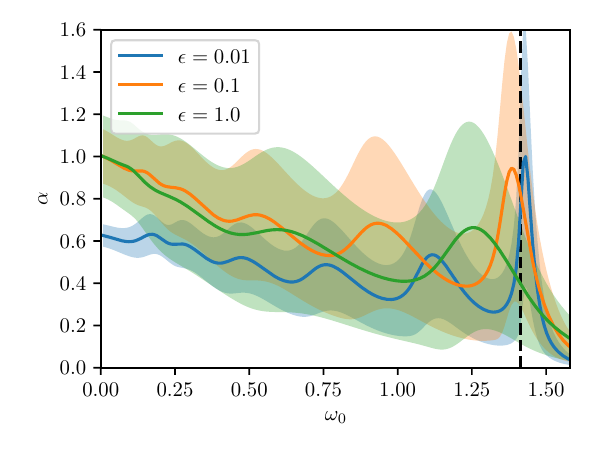}
			\caption{$ip$-method with $\boldsymbol{\Sigma} = \I$}
			\label{fig:cutoff_omega0_vs_alpha_ip_unweighted}
		\end{subfigure}
		\begin{subfigure}[b]{0.49\textwidth}
			\includegraphics[width=\textwidth]{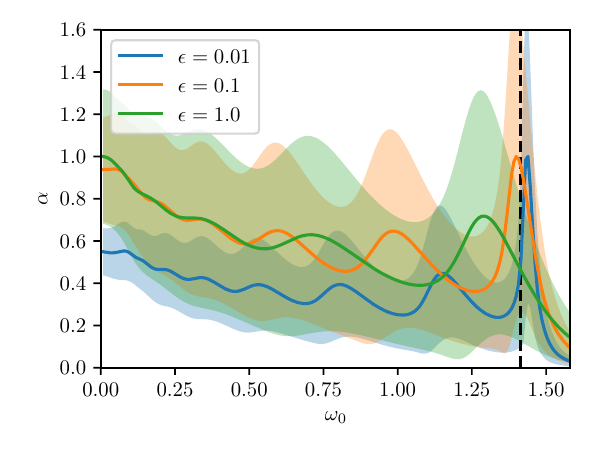}
			\caption{$p^2$-method with $\boldsymbol{\Sigma} = \I$}
			\label{fig:cutoff_omega0_vs_alpha_p2_unweighted}
		\end{subfigure}
		\caption{$\omega_0$ vs.~$\alpha$ for the cutoff model at $\Nres=128$.}
		\label{fig:cutoff_omega0_vs_alpha}
	\end{figure*}

	The optimal $\omega_0$ as a function of $\alpha$ for $\Nres = 128$ can be estimated from Figure \ref{fig:cutoff_omega0_vs_alpha}. Indeed, at low noise, $\alpha(\omega_0)$ is varying rapidly around the sharpest maximum which is located close to the physical cutoff $\omega_0^*$. This is also evident from the curve having the biggest standard deviation there, though it might not be immediately apparent from this plot. Following our earlier discussion on minimal dependence of $\omega_0$ on $\alpha$, we will use the location of this sharpest maximum as an estimator of $\omega_0^*$. Such a maximum appears for both methods and is more clearly present for the less noisy samples.

	Looking back at the curves $\alpha ( \omega_0 )$ for the Breit-Wigner and Bessel spectral functions in Figures~\ref{fig:bw_omega0_vs_alpha} and \ref{fig:bessel_omega0_vs_alpha}, we see that the last local maximum in the $\alpha ( \omega_0 )$ curve is the one with the largest variance of $\alpha$ w.r.t.~$\omega_0$ for all toy models studied, and therefore we infer that the last maximum in the $\alpha ( \omega_0 )$ curve provides best guess for the values of $\omega_0$ and $\alpha$.

\begin{figure*}[tbh]
		\centering
		\begin{subfigure}[b]{0.49\textwidth}
			\includegraphics[width=\textwidth]{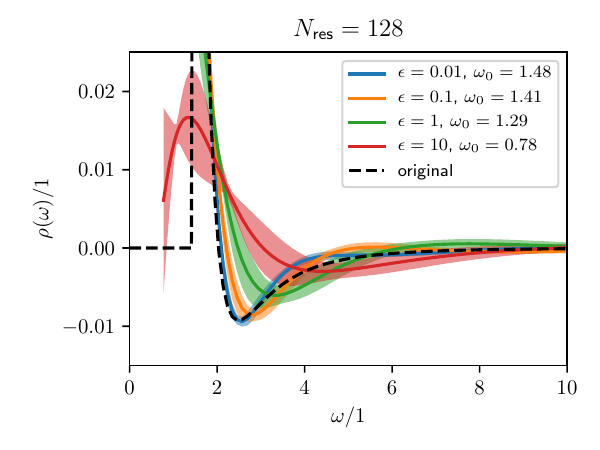}
			\caption{$ip$-method with $\omega_0 > 0$.}
			\label{fig:cutoff_omega0_vs_noomega0_ip}
		\end{subfigure}
		\begin{subfigure}[b]{0.49\textwidth}
			\includegraphics[width=\textwidth]{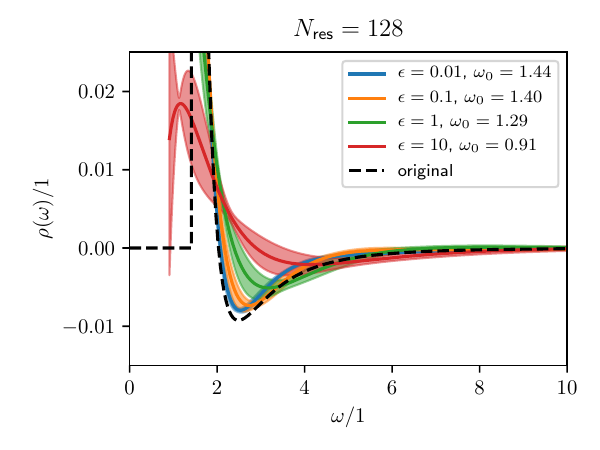}
			\caption{$p^2$-method with $\omega_0 > 0$.}
			\label{fig:cutoff_omega0_vs_noomega0_p2}
		\end{subfigure}
		\begin{subfigure}[b]{0.49\textwidth}
			\includegraphics[width=\textwidth]{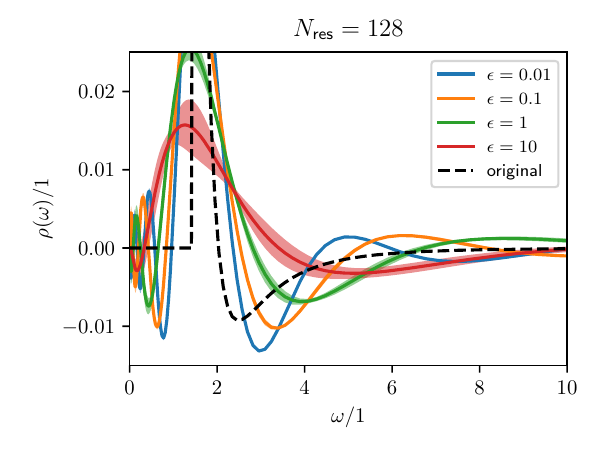}
			\caption{$ip$-method, $\omega_0 = 0$}
			\label{fig:cutoff_omega0_vs_noomega0_ip_unweighted}
		\end{subfigure}
		\begin{subfigure}[b]{0.49\textwidth}
			\includegraphics[width=\textwidth]{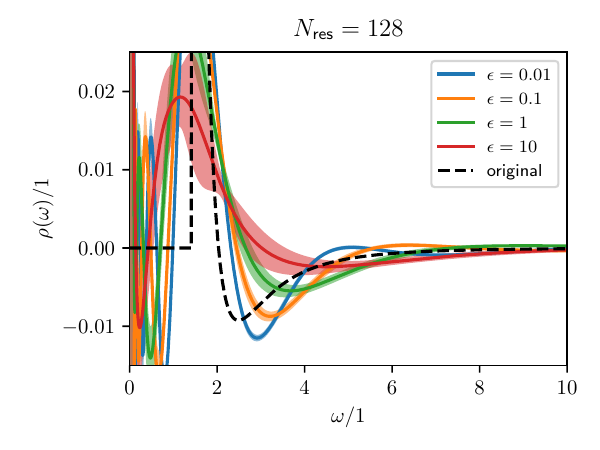}
			\caption{$p^2$-method, $\omega_0 = 0$}
			\label{fig:cutoff_omega0_vs_noomega0_p2_unweighted}
		\end{subfigure}
		\caption{Spectral density for the cutoff-model. All reconstructions used a $\boldsymbol{\Sigma} \ne \I$ and for $\omega_0 > 0$ its values were
			determined by the position of the last maximum in the corresponding Figure~\ref{fig:cutoff_omega0_vs_alpha}.}
		\label{fig:cutoff_omega0_vs_noomega0}
	\end{figure*}

	A comparison of the reconstructed spectral function taking $\alpha$ at the sharpest extremum of the curve $\alpha ( \omega_0 )$, corresponding to an $\omega_0 \simeq \sqrt{2}$,
	with setting $\omega_0 = 0$ can be seen in Figure~\ref{fig:cutoff_omega0_vs_noomega0} for both the $ip$- and $p^2$-method. In all these cases $\boldsymbol{\Sigma}$ was included.
	In general, for both methods the reconstructed spectral functions are quite similar outside of the IR and, clearly, the introduction of a finite cutoff gives a $\rho ( \omega )$ that is closer
	to the original input function. Moreover, both methods are sensitive to the maximum of the spectral function even at relatively large noise levels.

	%================================================================
	\subsection{Comparing the \texorpdfstring{$ip$}{ip} vs.~\texorpdfstring{$p^2$}{p2} method: numerical and analytical insights}
	\label{sec:p2_vs_ip}

	The analysis of the toy models suggests that, in general, the $ip$-method outperforms the $p^2$-method. In particular, the reconstructed spectral function shows a highly oscillatory
	behaviour in the IR for the $p^2$-method which is not observed with the $i p$-method; see e.g. Figures~\ref{fig:bw_fixed_epsilon} and \ref{fig:bw_fixed_Nres}.
	Moreover,  the $i p$-method has an overall larger $R^2$ as shown in e.g.~Figure~\ref{fig:bw_R2}.

	A first hint at this difference is given by the object $\boldsymbol{M}$ for both methods. Comparing Eq.~\eqref{eq:M_ip} with \eqref{eq:M_p2} and setting $\omega_0 = 0$, we find
	\begin{eqnarray}
		M_{ij} &=&
		\begin{cases}
			\frac{1}{p_j^2 - p_i^2}\ln(\frac{p_j^2}{p_i^2})  & i \neq j \\
			& \\
			\frac{1}{p_i^2}  & i = j
		\end{cases}
		\label{eq:M_p2_0}
	\end{eqnarray}
	for the $p^2$-method and
	\begin{eqnarray}
		M_{ij}&=&
		\begin{cases}
			\frac{2 \pi}{\abs{p_i} + \abs{p_j}} & p_i p_j \leq 0 \\
			& \\
			0 & \text{otherwise}
		\end{cases}
		\label{eq:M_ip_0}
	\end{eqnarray}
	for the $ip$-method.
	The latter is identical to that of a Laplace transform apart from the fact that it is only valid when $p_i p_j \leq 0$.

	In fact, it can be shown that the $p^2$-formalism can be directly obtained by performing $G = \Lp \Lp \rho$,
	\begin{align}
		\Lp_{t} \{ \Lp_{\mu} \{\rho(\sqrt{\mu})\}(t) \}(k) &= \int_{0}^{\infty} \dd{t} e^{- k^2 t} \pqty{\int_{0}^{\infty}  e^{- t \mu} \rho(\sqrt{\mu}) \dd{\mu}}  \nonumber\\
		&= \int_{0}^{\infty} \dd{t} e^{- k^2 t} \pqty{\int_{0}^{\infty} 2 \omega e^{- t \, \omega^2} \rho(\omega) \dd{\omega}} \nonumber\\
		&= \int_{0}^{\infty} \dd{\omega} \frac{2 \omega \rho(\omega)}{\omega^2 + k^2},
	\end{align}
	whereas the $ip$-formalism is obtained by performing $G = i \Lp \F \rho$:
	\begin{align}
		i \Lp_{t} \{ \F_{\omega} \{ \rho(\omega) \} (t) \} (k) &= i \int_{0}^{\infty} \dd{t} e^{- k t} \pqty{\int_{-\infty}^{\infty} \dd{\omega} e^{- i t \omega} \rho(\omega)} \nonumber\\
		&= i \int_{0}^{\infty} \dd{t} \int_{-\infty}^{\infty} \dd{\omega} e^{- (k + i\omega)t} \rho(\omega) \nonumber\\
		&= \int_{-\infty}^{\infty} \dd{\omega} \frac{\rho(\omega)}{\omega - i k}.
	\end{align}
	This helps to understand the difference in performance of the two methods. The Laplace transform is a common example of an ill-conditioned inversion problem,
	so doing it twice is not likely to improve the situation. On the other hand, the Fourier transform has a well-defined inversion and therefore this operation will not negatively affect the quality of the inversion.

\begin{figure}[htb]
		\centering
		\begin{subfigure}[b]{0.49\textwidth}
			\includegraphics[width=\textwidth]{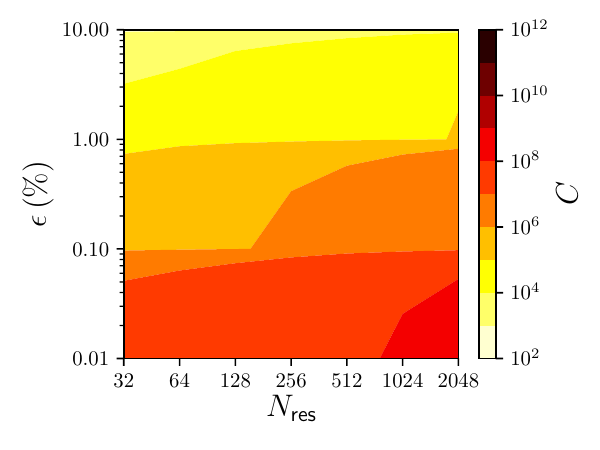}
			\caption{Condition number for the $ip$-formalism}
			\label{fig:bw_Cno_heatmap_repeat}
		\end{subfigure}
		\begin{subfigure}[b]{0.49\textwidth}
			\includegraphics[width=\textwidth]{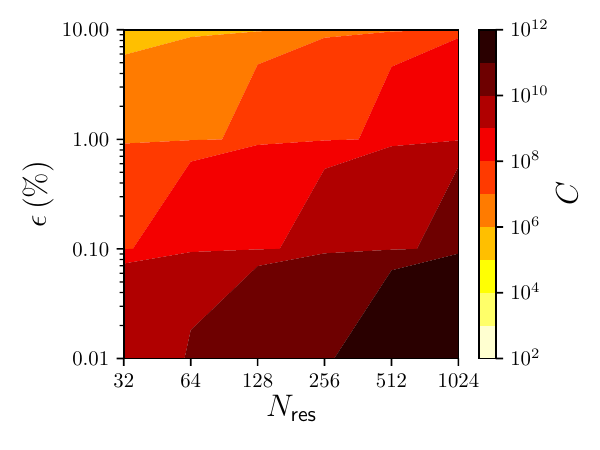}
			\caption{Condition number for the $p^2$-formalism}
			\label{fig:bw_p2_Cno_heatmap}
		\end{subfigure}
		\caption{Heatmaps of the condition number of $\pqty{\I + \frac{1}{\alpha^2} \boldsymbol{M} \boldsymbol{\Sigma}^{-1}}$ at median $\alpha$.}
		\label{fig:Cno_ip_vs_p2}
	\end{figure}

	The analysis of the condition number, defined via the ratio of the maximal and minimal singular value of the matrix $\pqty{\I + \frac{1}{\alpha^2} \boldsymbol{M} \boldsymbol{\Sigma}^{-1}}$, also helps understanding the difference between
	the two methods. Recall that this is the matrix which has to be inverted in order to find the residual $\boldsymbol{c}$.
	We have computed the condition number associated with the matrix for both methods at median $\alpha$ for the whole $(\epsilon, \Nres)$-range for the Breit-Wigner model.
	The results of this analysis are shown in Figure \ref{fig:Cno_ip_vs_p2} and it is apparent that the condition number of the $p^2$-algorithm is consistently 2--3 orders of
	magnitude larger, which could explain the different behaviour in the IR region.
	Also, the $i p$-method consistently reaches higher values of $R^2$ at lower statistical noise levels when compared to the $p^2$-method.

	Of the two methods considered herein, one can claim that overall the $ip$-method performs better and, therefore, for the
	analysis of the lattice data for the gluon and ghost propagators we will report only the results from this method. For the record, a $p^2$-method analysis of similar gluon and/or ghost data can be found in earlier work \cite{Dudal:2013yva,Oliveira:2016stx}.

	%============================================================================
	\section{The Landau gauge spectral functions from lattice data}
	\label{sec:gluon_and_ghost}

	We now proceed to compute the spectral function from the Landau gauge lattice gluon and ghost propagators  at $T = 0$.
	Due to rotational invariance at $T=0$, it is permissible to switch between the use of $p_4^2$ to $p^2$ as the fundamental variable in Eqs. (\ref{eq:kl_rep_p2}, \ref{eq:kl_rep_ip}).
	Indeed, as is well-known from e.g. \cite{Peskin:1995ev,Coleman:2011xi}, the standard variable in the K\"allén-Lehmann representation at $T=0$ is $p^2$.
	However, due to lattice effects this rotational invariance is violated, and a significant difference between $p_4^2$ and $p^2$ appears.
	In order to correct for this we have followed the standard technique of applying momentum cuts, as developed in the seminal papers \cite{Leinweber:1998im, Leinweber:1998uu} to deal with the breaking of rotational invariance.
	After these cuts the corrected $p^2$ provides the better measure, and is therefore used instead of the on-axis $p_4^2$. In Appendix \ref{app:cuts} we provide a comparison of these momentum sets to illustrate this point in detail.

	The lattice data for the gluon propagator was taken from~\cite{Dudal:2018cli}. For the ghost propagator we use the data published in~\cite{PhysRevD.94.014502}.
	The propagators were obtained from simulations on an $80^4$ lattice with $\beta = 6.0$, with lattice spacing $a = 0.1016(25)$ fm, corresponding to a physical volume of (8.1 fm)$^4$. The lattice data shown below refers to renormalized data within the MOM scheme at the scale $\mu = 4$ GeV, i.e.
	the scalar form factors associated with the gluon and ghost propagators are such that
	\begin{equation}
		\left.    G(p^2) \right|_{p^2 = \mu^2} = \frac{1}{\mu^2} \ .
	\end{equation}
	Details on the sampling, gauge fixing and definitions can be found in~\cite{Dudal:2018cli,PhysRevD.94.014502}.

	\begin{wrapfigure}{R}{0.5\textwidth}
		\centering
		\vspace{-20pt}
		\includegraphics[width=0.5\textwidth]{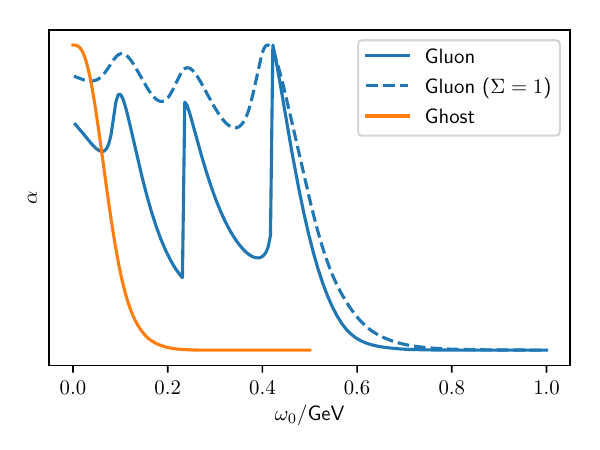}
		\caption{The regularization parameter $\alpha$ as a function of the IR cutoff $\omega_0$ for the reconstruction of ghost and gluon propagators.
			Note that the $\alpha$-axis is in arbitrary units.}
		\label{fig:omega0_vs_alpha}
	\end{wrapfigure}

	The lattice gluon propagator was computed with a large ensemble of 550 gauge configurations and its noise level is of the order
	$\epsilon \sim 0.5\%$ for the all $\Nres = 219$ momentum values.
	On the other hand, the ghost propagator was computed using a much smaller ensemble that included only 100 gauge configurations.
	However, the use of several sources considerably improves the quality of the ghost lattice data and the corresponding noise
	level is $\epsilon \sim 1\%$ or less for all the $\Nres = 219$ momentum values.
	In order to estimate the variance in the reconstructions we rely on the bootstrap method, where each bootstrap sample was inverted individually, giving an ensemble of spectral density functions which was then used to calculate the mean spectral density and its variance. In total 5500 bootstrap samples were considered for the gluons, and 700 for the ghosts. However, the reconstruction and spectral function of a bootstrap sample were only included in calculating the mean and standard deviation if the Morozov criterion was met to a precision of at least $10^{-10}$ during the inversion. This was true for about 25\% of the samples.

	From the results of the toy models for the noise levels and number of data points used in the inversion,
	see Figures~\ref{fig:bw_R2_rho_ip} and \ref{fig:bessel_R2_rho_ip}, one can expect an $R^2 \approx 0.9$ for the gluon inversion
	and an $R^2 \approx 0.8$ for the ghost inversion. Moreover, one also expects a good determination of the location and height of the absolute maximum of
	the gluon and ghost spectral functions.
	We call the readers attention to the fact that our analysis does not take into account possible systematics, nor do we account for correlations between the
	different momenta. From the technical point of view, this last remark means that we only consider variances, not covariances.
	This refinement, amongst other things, will be discussed in future work.

	In Figure~\ref{fig:omega0_vs_alpha} we report the curves $\alpha ( \omega_0 )$ for the inversion of the gluon data when $\Sigma_{ij} = \sigma^2_i \delta_{ij}$ (no sum) and $\Sigma_{ij} = \delta_{ij}$, i.e.~with or without taking into account the statistical errors during the inversion, and for the ghost inversion with $\Sigma_{ij} = \sigma^2_i \delta_{ij}$ (no sum).
	Taking $\Sigma_{ij} = \delta_{ij}$ is not displayed because for the ghosts this does not yield any interesting new information.
	The pattern of the $\alpha ( \omega_0 )$ curves for the gluon and ghost inversions look rather different.
	For the gluon inversion, $\alpha ( \omega_0 )$ has several extrema that can be associated with several values of the cutoff $\omega_0$
	where $\partial \omega_0 / \partial \alpha \simeq 0$. On the other hand, for the ghost inversion the $\alpha(\omega_0)$ curve has a single maximum at $\omega_0 = 0$
	and a steep decrease towards small values of $\alpha$ as $\omega_0$ departs from zero towards larger values.
	We take this behaviour as an indication that the right cutoff value for the ghost data is $\omega_0 = 0$ and hence we will only display the results for the ghost inversion at this particular cutoff.
	In the inversions of the gluons and ghosts the diagonal covariance matrix $\Sigma_{ij} = \sigma^2_i \delta_{ij}$ (no sum) was always included since our toy model studies showed that this gives better reconstructions than without including a covariance matrix.

	We do not attempt to check the sum rule \eqref{eq:sumrule} as in our formulation, by construction, the correct UV asymptotic logarithmic tails of neither propagator nor spectral function are reproduced.
	This is best seen from Eqs.~(12)-(16) in \cite{Dudal:2013yva}, and the discussion thereafter.
	Roughly speaking the current Tikhonov implementation gives, for $\mu$ large, $\rho(\mu) \sim1/\mu$ and $G(p)\sim (\ln p^2)/p^2$.
	As mentioned earlier, the situation could be improved by including the sum rule as a constraint.
	We will come back to this issue in future work.

	%==========================================================
	\subsection{The gluon spectral function}

	As can be seen in Figure~\ref{fig:omega0_vs_alpha}, for the gluon data inversion the curve $\alpha ( \omega_0 )$ shows several regions where $\alpha$ changes quickly as $\omega_0$ varies slightly, which can be associated with values of the cutoff $\omega_0$ that are stable against variation of the
	regularization parameter $\alpha$, i.e.~where $\partial \omega_0 / \partial \alpha \simeq 0$.
	The precise values where the derivative vanishes are more difficult to identify. From a practical point of view, and based on the observations for the toy models, we take the location of the
	corresponding nearby maxima as the estimated value for the IR cutoff. However, inversions are also performed at the locations of the minima in this curve to allow for a direct comparison.

\begin{figure*}[bth]
		\centering
		\begin{subfigure}[t]{0.47\textwidth}
			\begin{center}
				\includegraphics[width=\textwidth]{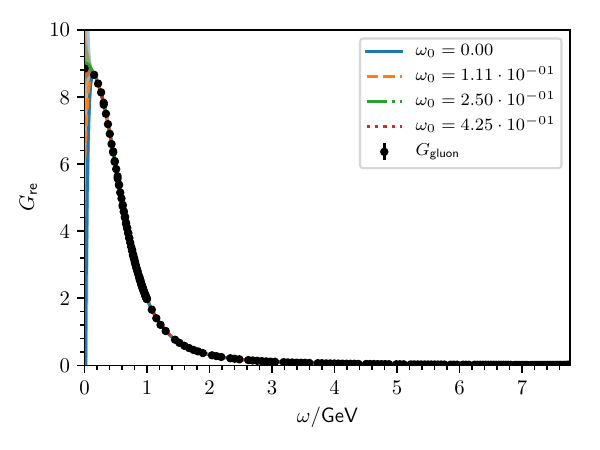}
				\caption{Reconstructed gluon propagator for the full range of lattice momenta for the $\omega_0$ associated with the maxima of $\alpha ( \omega_0 )$.}
				\label{fig:G_gluon}
			\end{center}
		\end{subfigure} ~ ~
		\begin{subfigure}[t]{0.47\textwidth}
			\begin{center}
				\includegraphics[width=\textwidth]{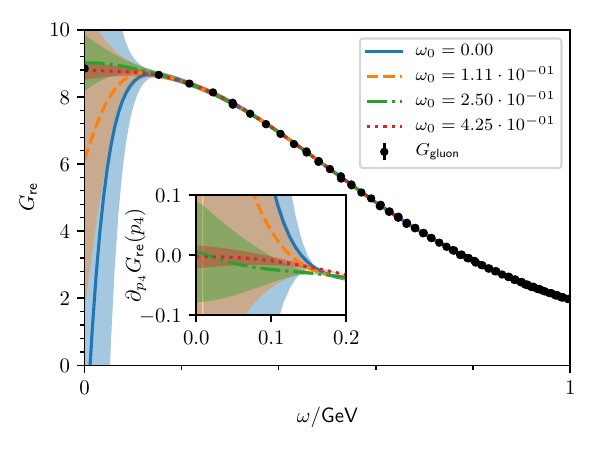}
				\caption{IR reconstructed propagator and its derivatives for the reconstructions associated with the maxima of $\alpha ( \omega_0 )$.}
				\label{fig:G_gluon_zoom}
			\end{center}
		\end{subfigure}
		\begin{subfigure}[t]{0.47\textwidth}
			\begin{center}
				\includegraphics[width=\textwidth]{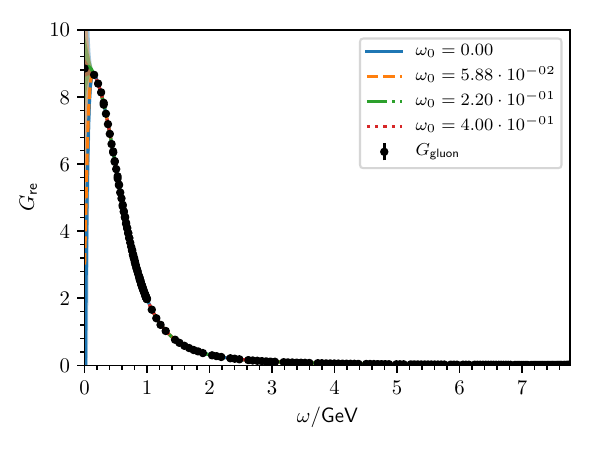}
				\caption{Reconstructed gluon propagator for the full range of lattice momenta for the $\omega_0$ associated with the minima of $\alpha ( \omega_0 )$.}
				\label{fig:G_gluon_min}
			\end{center}
		\end{subfigure} ~ ~
		\begin{subfigure}[t]{0.47\textwidth}
			\begin{center}
				\includegraphics[width=\textwidth]{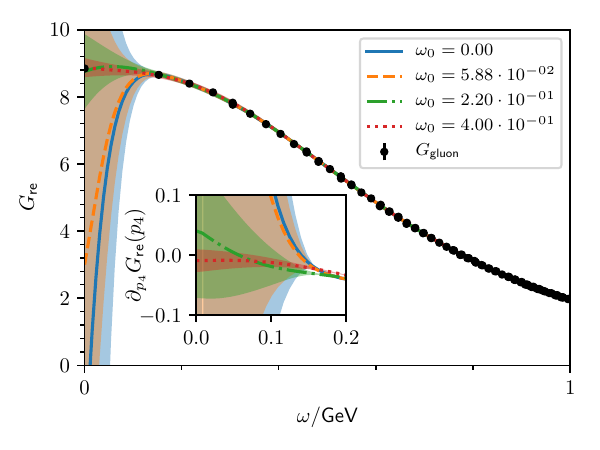}
				\caption{IR reconstructed propagator and its derivatives for the reconstructions associated with the minima of $\alpha ( \omega_0 )$.}
				\label{fig:G_gluon_min_zoom}
			\end{center}
		\end{subfigure}
		\caption{Reconstructed gluon propagator for all maxima (top) and minima (bottom) of the $\alpha ( \omega_0)$ curve.}
		\label{fig:gluon}
	\end{figure*}

	Figure~\ref{fig:gluon} shows the reconstructions of the gluon propagator for the $\omega_0$ identified with the extrema of $\alpha ( \omega_0 )$ as shown in Figure~\ref{fig:omega0_vs_alpha}. As for the toy models, the $p=0$ data point was not included in the inversion procedure. In general, the reconstructed propagators are in very good agreement with the lattice data for all $\omega_0$. However, when extrapolating towards $p \to 0^+$, it is clear that the introduction of a cutoff $\omega_0$ greatly improves the prediction of the point $G(p^2 = 0)$, with the minimum $\omega_0=400$~MeV and the maximum $\omega_0=425$~MeV performing most reliably. The minimum at $\omega_0=220$~MeV and the maximum $\omega_0=250$~MeV also still perform reasonably well, but the reconstructions with $\omega_0=0$~MeV,  $\omega_0=59$~MeV and $\omega_0=111$~MeV are most certainly unreliable.

\begin{figure*}[bth]
		\begin{subfigure}[t]{0.49\textwidth}
			\includegraphics[width=\textwidth]{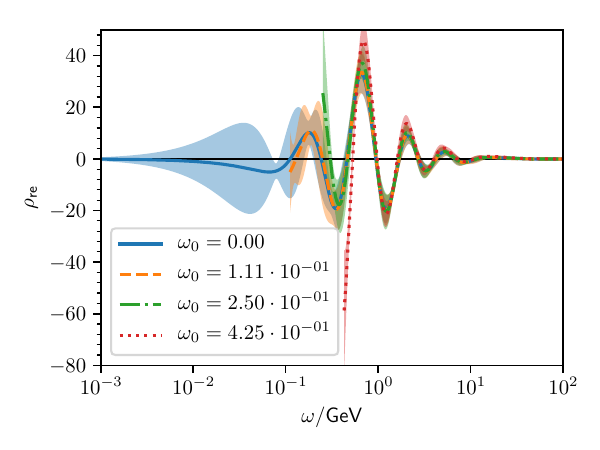}
			\caption{Spectral densities at the maxima of $\alpha(\omega_0)$, logarithmic scale.}
			\label{fig:rho_gluon_log}
		\end{subfigure}
		\begin{subfigure}[t]{0.49\textwidth}
			\includegraphics[width=\textwidth]{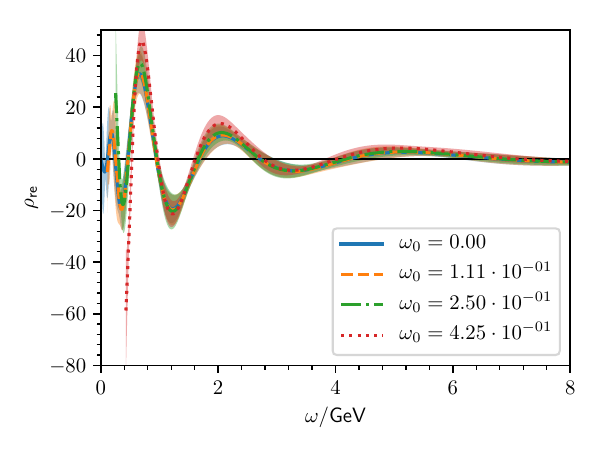}
			\caption{Spectral densities at the maxima of $\alpha(\omega_0)$, linear scale.}
			\label{fig:rho_gluon_lin}
		\end{subfigure}
		\begin{subfigure}[t]{0.49\textwidth}
			\includegraphics[width=\textwidth]{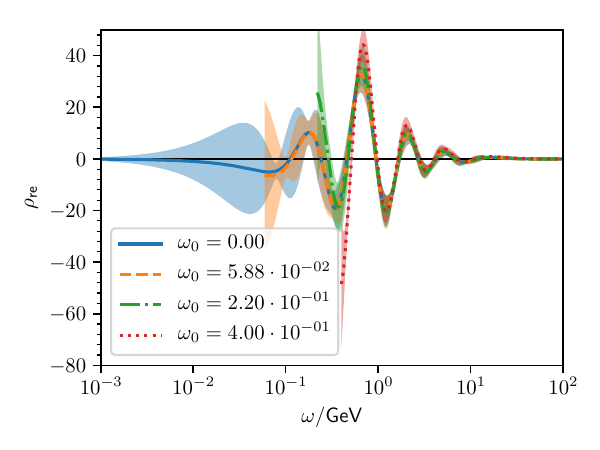}
			\caption{Spectral densities at the minima of $\alpha(\omega_0)$, logarithmic scale.}
			\label{fig:rho_gluon_min_log}
		\end{subfigure}
		\begin{subfigure}[t]{0.49\textwidth}
			\includegraphics[width=\textwidth]{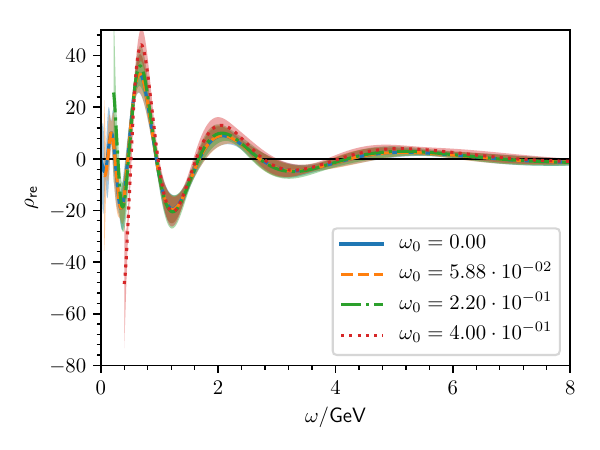}
			\caption{Spectral densities at the minima of $\alpha(\omega_0)$, linear scale.}
			\label{fig:rho_gluon_min_lin}
		\end{subfigure}
		\caption{Spectral densities for the reconstructions reported in Figure~\ref{fig:gluon}.}
		\label{fig:rho_gluon}
	\end{figure*}

	With this in mind we turn our attention to the spectral densities associated with the various extrema of $\alpha ( \omega_0 )$, as shown in Figure~\ref{fig:rho_gluon}.
	One striking observation is the stability of the positions of the zeroes and the extrema of $\rho ( \omega )$ for all reconstructions, even if the corresponding values are not exactly identical.
	The effect of reconstructing at a maximum or minimum of $\alpha(\omega_0)$ seems minimal on these features.
	It follows that one can claim a global maximum for the spectral function at $\omega = 0.65$ GeV with ``Full width at half maximum''(FWHM) = $0.27$ GeV, a negative minimum at $\omega = 1.19$ GeV with FWHM = $0.49$ GeV, a positive maximum at $\omega = 2.11$ GeV with FWHM = $0.71$ GeV, etc. with zeros of $\rho ( \omega )$ between the quoted $\omega$ values.

	The computed spectral functions reproduce the pattern observed in the preliminary study~\cite{Oliveira:2016stx}, where $\rho$ reached a maximum at momentum
	$\sim 0.5 - 0.6$ GeV and then oscillated, approaching zero at higher $\omega$. The herein computed absolute maximum of the spectral function
	is roughly consistent with the predictions of \cite{Cyrol:2018xeq} that used the numerical outcome from functional renormalization group (FRG) equations compatible with the scaling
	scenario, i.e.~a gluon propagator going to zero at zero momentum following a simple power law. The FRG spectral function was calculated using a Bayesian inspired approach
	that includes a dedicated guess for a basis of functions and takes into account \textit{a priori} knowledge about its asymptotic behaviour.
	Once more, the general pattern of the spectral function computed here is in qualitative agreement with that computed in \cite{Cyrol:2018xeq}, an absolute maximum followed by an oscillatory behaviour towards zero, although the quantitative details differ.
	It should be noted, however, that contrary to \cite{Cyrol:2018xeq}, our method makes no explicit assumptions about the IR of the spectral function, i.e.~we make no assumptions on the structure of the propagator in the IR\footnote{In the UV, model independent analytical estimates can be given for the spectral function based on the perturbative renormalization group, see also our Appendix~\ref{A}. In the IR, such estimates are usually model dependent. We refrained from building in the correct UV asymptotics via appropriate choice of a prior estimate for $\rho$ in the regulating part of the Tikhonov functional. Tests indicated that this has little to no influence on the reconstruction in the mid-momentum regime where most of the interesting phenomenology happens.}. The price paid is that we observe the oscillations (``ringing'') in the IR.

	It is known that different regularization recipes for ill-posed inversion problems, of which the Källén-Lehmann spectral integral calculation is an example, can yield different results. Therefore, it is important to have several toolkits to test the soundness of the computed spectral functions.
	This is a common observation, even applicable when gauge invariant lattice data are inverted for e.g. meson spectral functions \cite{Asakawa:2000tr}.
	But because the same dominant peak is found for all cutoffs $\omega_0$, as well as by \cite{Cyrol:2018xeq}, it is fair to say that this peak is meaningful.

	Given a functional form for the spectral function as in Eq.~(\ref{eq:tikhonov_vec_rho}), one can measure its derivatives. Indeed,
	it can also be shown~\cite{Cyrol:2018xeq} that in the limit of $p_4 \to 0^+$, $\partial_{p_4} G(p_4) = -\partial_{p_4} \rho(p_4) / 2$.
	The inset of Figure \ref{fig:G_gluon_zoom} shows the derivative of the various reconstructions for small momenta.
	For all the reconstructions where a cutoff $\omega_0 \geqslant 200$~MeV has been included, the derivatives go to zero within error, which is consistent with having $\rho(\omega) = 0$ below the cutoff, and hence $\partial_{\omega} \rho(\omega) = 0$, confirming the above result.
	As $\omega$ increases, this simple relationship between $\partial_{p_4} G(p_4)$ and $\partial_{p_4} \rho(p_4)$ breaks down, so no conclusions can be drawn other than that the $\omega \to 0$ behavior is correct for cutoff values $\omega_0 \geqslant 200$~MeV.
	For the inversions at cutoffs $\omega_0 < 200$~MeV, the relation between the derivatives of the propagator and of the spectral function is not satisfied, suggesting that a cutoff has to be included.

	%=======================================================================
	%=======================================================================
	\subsection{Ghost propagator}

	As discussed previously, for the ghost propagator there is no ambiguity regarding the choice of $\omega_0 = 0$.
	The ghost propagator is expected to be massless and, therefore, its spectral function should have a Dirac delta function for zero momentum.
	But if this is the case, the inversion of the ghost propagator data becomes rather difficult, if not impossible, to perform. Alternatively, one can rely on the ghost dressing function
	given by
	\begin{align}
		g(p) = p^2 G(p) = \int_{-\infty}^{\infty} \frac{\sigma(\omega)}{\omega - i p} \dd{\omega},
		\label{eq:ghost_dress}
	\end{align}
	where $\sigma(\omega)$ is the corresponding spectral density function. Introducing the function
	\begin{equation}
		\hat{\rho}(\omega) \coloneqq - \frac{\sigma(\omega)}{\omega^2}
		\label{eq:rho_hat}
	\end{equation}
	it follows, after integrating $\hat{\rho}$, that
	\begin{equation}
		\hat{G}(p) \coloneqq \int_{-\infty}^{\infty} \frac{\hat{\rho}(\omega)}{\omega - i p} \dd{\omega} = - \frac{g(0)}{p^2} + G(p)
		\label{eq:ghost_dress_prop}
	\end{equation}
	and $\hat{G}(p)$ equals $G(p)$ up to the term $- g(0)/p^2$. In order to cancel this additional term, $\rho(\omega)$ has to be given by
	\begin{equation}
		\rho(\omega) = - g(0) \delta'(\omega) + \hat{\rho}(\omega),
		\label{eq:rho_ansatz}
	\end{equation}
	which can be checked by plugging Eq.~\eqref{eq:rho_ansatz} into Eq.~\eqref{eq:kl_rep_ip} and performing the integral.
	A more detailed derivation of Eq.~\eqref{eq:rho_ansatz} can be found in Appendix~\ref{app:rho_ghost}. The significance of Eq.~\eqref{eq:rho_ansatz} is the following: by inverting $g(p)$ instead, $\hat{\rho}$ can be built. This is identical to $\rho$, apart from the fact that an additional $\delta'(\omega)$ has to be present at the origin. Keeping this $\delta'(\omega)$ in mind, we can therefore consider $\hat{\rho}$ as the spectral function after subtracting the massless free ghost state.

	\begin{figure*}
		\centering
		\begin{subfigure}[t]{0.48\textwidth}
			\includegraphics[width=\textwidth]{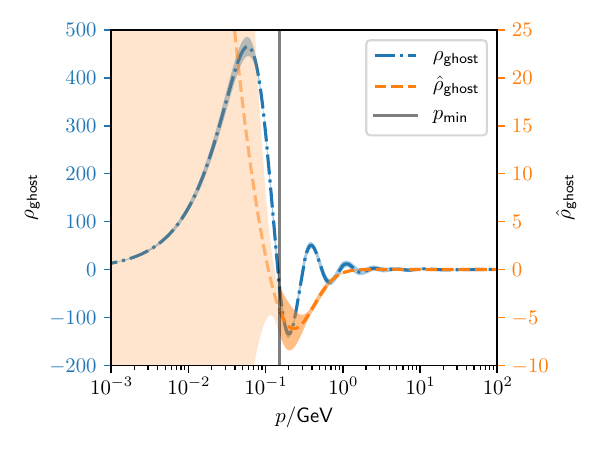}
			\caption{Spectral density before ($\rho_\text{ghost}$) and after $\delta (0)$ removal ($\hat{\rho}_\text{ghost}$). The vertical black line indicates $p_\text{min} = 0.15$ GeV, the smallest momentum value measured.}
			\label{fig:rho_ghost_from_dressing_function}
		\end{subfigure} ~ ~
		\begin{subfigure}[t]{0.48\textwidth}
			\includegraphics[width=\textwidth]{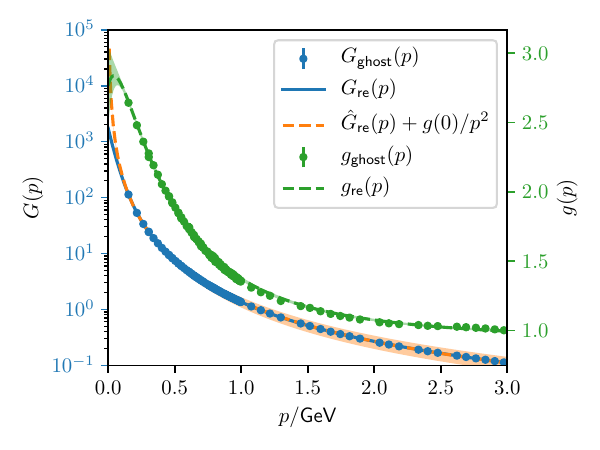}
			\caption{Reconstruction of the ghost propagator and dressing function. Error bars are not significant compared to the resolution of the plot.}
			\label{fig:g_ghost}
		\end{subfigure}
		\caption{The ghost spectral function (left), and reconstructed propagator and dressing function (right).}
		\label{fig:ghost_delta_removal}
	\end{figure*}

	\begin{wrapfigure}{R}{0.5\textwidth}
		\centering
		\vspace{-20pt}
		\includegraphics[width=0.5\textwidth]{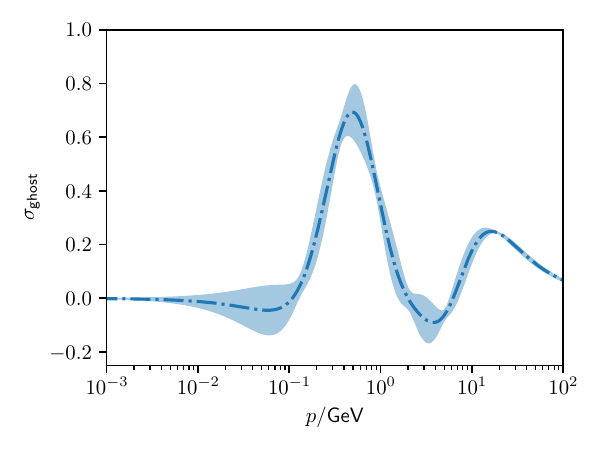}
		\caption{The inverse of $g(p)$, $\sigma(\omega)$.}
		\label{fig:sigma_ghost}
	\end{wrapfigure}
		Figure \ref{fig:rho_ghost_from_dressing_function} presents both $\rho$ as obtained from a direct inversion of $G(p)$, and $\hat{\rho}$ as defined previously.
		Both reconstructions display the same minimum at $p\approx 200$~MeV, but $\hat{\rho}$ starts to be dominated by the $1/\omega^2$ behaviour for smaller momenta whereas $\rho$ shows a maximum at $p \approx 70$~MeV before going down to zero.
		The vertical black line in \ref{fig:rho_ghost_from_dressing_function} indicates the smallest momentum value in the dataset, $p_\text{min}=0.15$ GeV.
		The maximum of $\rho$ is positioned below $p_\text{min}$, indicating that this peak could possibly be due to the attempt to reconstruct the Dirac-$\delta$ peak at $p=0$.
		The error in $\hat{\rho}$ explodes below $p_\text{min}$, despite the fact that the error in $\sigma(\omega)$ stays reasonable, as can be seen from Fig. \ref{fig:sigma_ghost}. This is a direct consequence of the $1 / \omega^2$ behaviour of $\hat{\rho}$, however the typical solution does look like the average shown in Fig. \ref{fig:rho_ghost_from_dressing_function}.

	To test our approach, it has been verified that $\hat{G}(p) + g(0)/p^2$ is also a reconstruction of the original lattice data for the propagator, which is indistinguishable from
	a direct reconstruction of $G(p)$. This can be seen in Figure \ref{fig:g_ghost}, which shows the reconstructed propagator, and clearly demonstrates that $\hat{G}(p) + g(0)/p^2$
	overlays the lattice data over the full $p$-range where data points have been provided.
	To reconstruct the ghost propagator in this manner, $\hat{\rho}$ was integrated according to Eq. \eqref{eq:ghost_dress_prop} to yield $\hat{G}(p)$, and $g(0)$ was calculated using Eq. \eqref{eq:ghost_dress}.
	The good agreement found is an indication that the infrared ghost propagator is given essentially by its tree level value.

	\section{Conclusion}
	In the current paper, we improved the Tikhonov reconstruction using the Morozov discrepancy principle when applied to Källén-Lehmann inverse lattice spectroscopy, as set out previously in \cite{Dudal:2013yva}.
	Based on dedicated toy models, the current research defined statistical measures for the quality of the inversion, providing a region of validity for this method.
	While doing so, we also supplemented the Tikhonov functional with a relative error weighting of the data.

	We considered two analytically equivalent versions of the Källén-Lehmann spectral representation.
	However, despite this analytical equivalence, the numerical performance of the two formalisms is quite different.
	This difference is most notable in the IR, where the $ip$-formalism yields significant improvement over the previous $p^2$-formalism \cite{Dudal:2013yva}.
	The improvement was demonstrated most notably by a reduction in the condition number of the to-be-inverted matrix
	$\pqty{\I + \frac{1}{\alpha^2} \boldsymbol{M} \boldsymbol{\Sigma}^{-1}}$ by 2--3 orders of magnitude.
	Both methods are sensitive to an IR-cutoff below which the spectral density vanishes, if such a cutoff is present.

	By applying the $ip$-methodology to lattice SU(3) gluon data, it was found that the gluon spectral function seems to have an IR cutoff of a few hundred MeV.
	We also characterized a dominant peak, the location of which is fairly consistent with the findings reported in other studies like \cite{Dudal:2013yva,Cyrol:2018xeq,Strauss:2012dg}. Because IR oscillations are always present in the reconstructions, we have to be careful with their interpretation. However, the dominant peak appears to be a stable prediction.
	For ghost data it was found that no IR-cutoff is present, consistent with the massless pole present in the ghost propagator.
	However, the quality of the reconstruction was greatly improved by first removing the $\delta$-peak in the spectral density.
	We stress once more that our results should always be considered under the \emph{assumption} that the gluon and ghost degrees of freedom have a Källén-Lehmann spectral representation to begin with.
	The existence of such a representation is not self-evident, since these particles are confined at zero temperature, and therefore they do not belong to the complete set of (positive norm) physical states that are usually employed to derive the spectral representation \cite{Negele:1988vy,Peskin:1995ev,Laine:2016hma}.
	Other analytical continuations of the data are thus in principle possible.
	We recall that the Källén-Lehmann spectral integral allows only for branch cuts along the negative real axis, while certain analytical approaches, or fits to lattice data based thereon, entail the presence of e.g.~complex conjugate poles, in se incompatible with the Källén-Lehmann spectral structure \cite{Cyrol:2018xeq,Baulieu:2009ha,Dudal:2010cd,Cucchieri:2011ig,Dudal:2018cli,Siringo:2016jrc,Siringo:2017svp}.
	As a future improvement of our method, such complex conjugate poles could therefore be included in the inversion method. The inclusion of such poles could also lead to a reduction in the IR ringing, which would be an indication of their presence. This being said, it should also be noted that the ringing effect is also present in the presented toy model inversions, which definitely contain no complex conjugate poles'~contributions. Moreover, the IR ringing also plagues other inversion strategies, like MEM \cite{Asakawa:2000tr,Tripolt:2018xeo}, again without such poles. In recent work, \cite{Binosi:2019ecz}, yet another inversion method was proposed, based on rational function interpolation. More evidence was presented for a single set of complex conjugate poles, so it would be interesting to test whether this feature prevails also within our methodology. This is currently under investigation.
	Additionally, imposing adherence to the sum rule as a constraint seems compatible with the $ip$~method and will be investigated further. Such a constraint could be extended to the generalized sum rule that includes the set of complex conjugate poles, see also the recent paper \cite{Hayashi:2018giz}.

	Moreover, in this next phase of research we will further put our inversion strategy to the test by applying it to finite temperature lattice data for gluons, ghosts and quarks.
	An important question to be further addressed there is whether, and to what extent if so, the spectral functions are sensitive to the deconfinement (chiral) transition or which kind of quasi-particle behaviour can be identified \cite{Qin:2013ufa,Ilgenfritz:2017kkp,Maas:2011se}.
	This will be discussed in forthcoming work.

	\section*{Acknowledgments}
	We benefitted from discussions with J.~Pawlowski and N.~Wink. The research of D.D.~and M.R.~is supported by KU Leuven IF project C14/16/067. D.D~and M.R.~are grateful for the hospitality and support from the University of Coimbra, whilst likewise the authors O.O.~and P.J.S.~are grateful for the hospitality and support from the KU Leuven, campus Kortrijk.
	The authors O.O.~and P.J.S.~acknowledge the Laboratory for Advanced Computing
	at University of Coimbra (\href{http://www.uc.pt/lca}{http://www.uc.pt/lca}) for providing access to the HPC computing resource
	Navigator. P.J.S.~acknowledges support by FCT under contracts
	SFRH/BPD/40998/2007 and SFRH/BPD/109971/2015. The SU(3) simulations were done using Chroma \cite{Edwards:2004sx} and
	PFFT \cite{pfft2013} libraries.

	\appendix
	\section{UV asymptotics of the spectral function }\label{A}
	\subsection{Leading log resummation of the propagator}
	Consider a propagator $G(p^2)=\left\langle O(p)O(-p)\right\rangle$, which we renormalize in a MOM (``momentum subtraction'') scheme at $p^2=\mu^2$, that is $G(\mu^2)=\frac{1}{\mu^2}$. Such a scheme can also be implemented on the lattice. As is well-known, we can resum the leading logs using the one-loop renormalization group equation, which leads to
	\begin{equation}\label{rg8}
		G(p^2)= \frac{1}{p^2}\left(1+\beta_0g^2\ln\frac{p^2}{\mu^2}\right)^{\frac{\gamma_0}{\beta_0}}.
	\end{equation}
	We used the conventions that
	\begin{equation}\label{rg3}
		\mu\frac{\p}{\p \mu} g^2=\beta(g^2)=-2\beta_0g^4+\ldots\,,\qquad \mu\frac{\p}{\p \mu} O=\gamma(g^2)O=\gamma_0g^2O+\ldots.
	\end{equation}
	In pure gauge theory with $N$ colors, one has (see e.g.~\cite{Chetyrkin:2000dq,Gracey:2013sca})
	\begin{equation}\label{rg6}
		\beta_0=\frac{11}{3}\frac{N}{16\pi^2}\,,\quad \gamma_0^{gl}=-\frac{13}{6}\frac{N}{16\pi^2}\,,\quad \gamma_0^{gh}=-\frac{3}{4}\frac{N}{16\pi^2}.
	\end{equation}

	\subsection{The UV spectral density after a leading log resummation}
	For sufficiently large $p^2$, the RG resummed propagator \eqref{rg8} will give a decent description of the lattice data.  We thus consider the expression \eqref{rg8} and wonder what the underlying spectral function would be, taking $p^2$ sufficiently large w.r.t.~$\mu$.  Since from a Källén-Lehmann representation in the form \eqref{eq:kl_rep_p2_mu}, we have (see for example (3.19) in \cite{Dudal:2010wn}):
	\begin{equation}
		\tilde\rho(t)=\frac{1}{2\pi i}\lim_{\epsilon\to 0^+}\left[G(-t-i\epsilon)-G(-t+i\epsilon)\right],
	\end{equation}
	we find for $t$ sufficiently large, with $\gamma=-\frac{\gamma_0^{gl,gh}}{\beta_0}$,
	\begin{eqnarray}
		\tilde\rho_{\gg}(t) &=& \frac{1}{2\pi i}\left[\frac{\left(\beta_0g^2\ln\frac{-t-i\epsilon}{\mu^2}+1\right)^{-\gamma}}{-t-i\epsilon}-\frac{\left(\beta_0g^2\ln\frac{-t+i\epsilon}{\mu^2}+1\right)^{-\gamma}}{-t+i\epsilon}\right]\nonumber \\
		&=&\frac{1}{2\pi i t} \left[-\left(\beta_0g^2\ln\frac{t}{\mu^2}-i\beta_0g^2\pi+1\right)^{-\gamma}+\left(\beta_0g^2\ln\frac{t}{\mu^2}+i\beta_0g^2\pi+1\right)^{-\gamma}\right]\nonumber\\
		&=&\frac{1}{\pi t}\text{Im}\left[\left(\beta_0g^2\ln\frac{t}{\mu^2}+i\beta_0g^2\pi+1\right)^{-\gamma}\right]\nonumber\\
		&=&\frac{1}{\pi t}\left(\left(\beta_0g^2\ln\frac{t}{\mu^2}+1\right)^2+\beta_0^2g^4\pi^2\right)^{-\gamma/2}\sin\left(-\gamma\arctan\frac{\beta_0g^2\pi}{\beta_0g^2\ln\frac{t}{\mu^2}+1}\right).
	\end{eqnarray}
	Notice that the spectral integral of the foregoing expression will \emph{not} be $G(p^2)$, given that the full $\rho(t)$ is different from $\tilde\rho_\gg(t)$. Indeed, Eq.~\eqref{rg8} also displays a cut for $p^2>0$ sufficiently close to zero in which case the $\ln$ will overtake the $+1$. Clearly, this will not contribute to $\tilde\rho_\gg(t)$ if $t$ is sufficiently large. We also notice that $\tilde\rho_\gg(t)$ becomes negative for $t$ large\footnote{This is known, see e.g.~\cite{Oehme:1990kd}.}.

	At lowest order, we can write
	\begin{eqnarray}
		\tilde\rho_{\gg}(t) &\stackrel{t\to\infty}{\sim}& -(\beta_0g^2)^{-\gamma}\frac{\gamma  }{t}\left(\ln\frac{t}{\mu^2}\right)^{-\gamma-1}. \label{rhoas}
	\end{eqnarray}
	We notice the foregoing result clearly dictates the spectral function of the gluon and ghost propagator to become negative at sufficiently large values of $t$.
	\subsection{Corollary: a sum rule}
	From the asymptotic behaviour \eqref{rg8}, we infer that both gluon and ghost propagator fall off faster than $\frac{1}{p^2}$, from which it follows from Eq.~\eqref{eq:kl_rep_p2} in the $p_4^2\to \infty$ limit that
	\begin{equation}\label{sumrulebis}
		\int_{\omega_0}^{\infty} \rho(\omega) \omega \dd{\omega}=0.\end{equation}
	This relation is also known as a superconvergence relation \cite{Oehme:1990kd}. From this relation, it is also evident that $\rho(\omega)$ cannot be positive-definite.

	\section{Gluon propagator rotational invariance breaking}\label{app:cuts}
\begin{figure*}[t]
		\includegraphics[width=\textwidth]{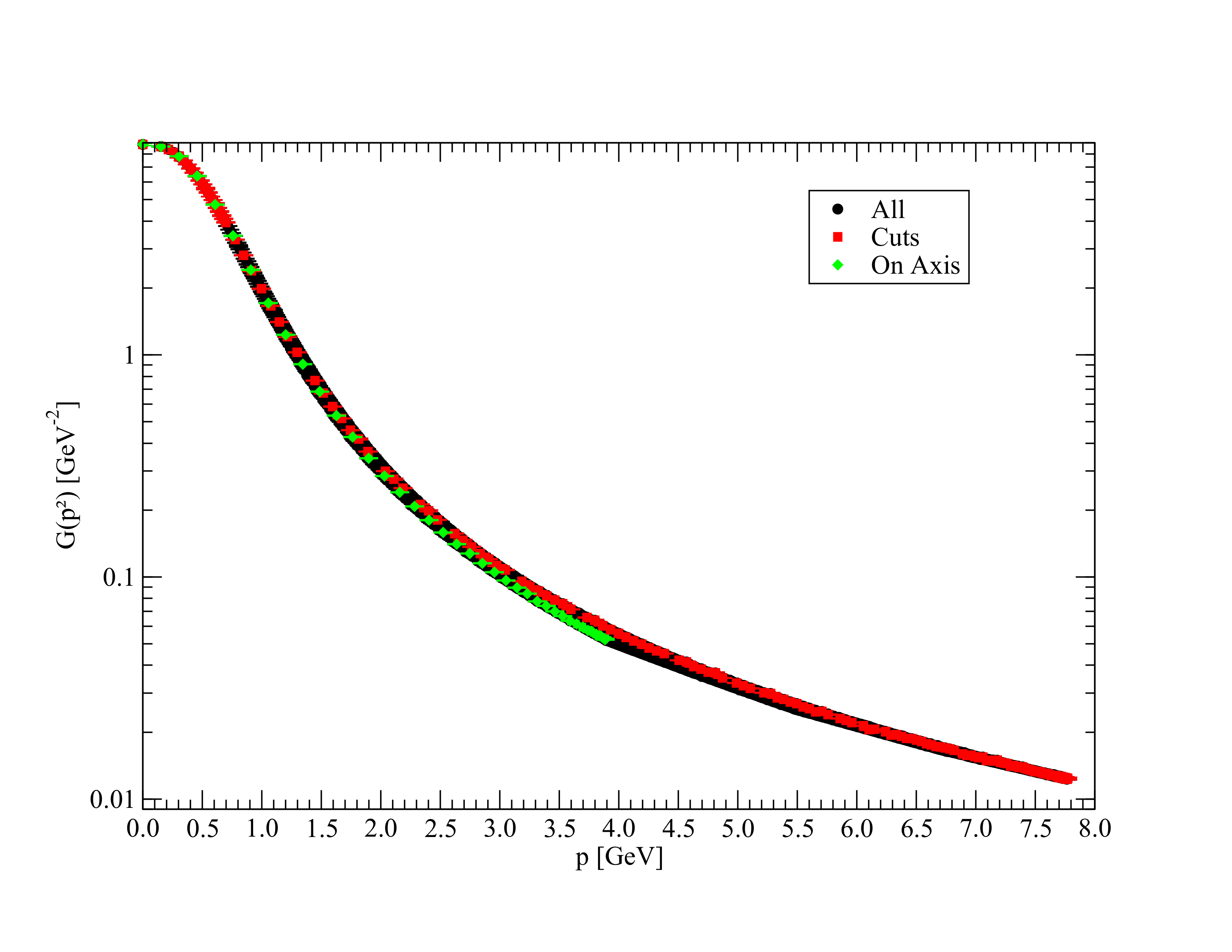}
		\caption{Gluon propagator for different types of momenta: all momentum values (black), momentum values obtained by performing cuts i.e.~close to the diagonal $p_\mu = (1, 1, 1, 1)$ (red), and  the on-axis values (green).}
		\label{fig:cuts}
	\end{figure*}

	In Figure~\ref{fig:cuts} we illustrate the effects associated with the breaking of rotational invariance in lattice simulations by comparing the gluon propagator lattice data for different types of momenta.
	In order to do so Figure~\ref{fig:cuts} displays the gluon propagator $G(p^2)$ at all momenta, the on-axis momenta, and the momentum cuts which are close to $p_\mu = (1, 1, 1, 1)$.
	These momentum cuts \cite{Leinweber:1998im, Leinweber:1998uu} were devised to suppress rotational breaking effects on the propagator lattice data, and as an additional advantage they also provide access to a larger range of momenta when compared to the on-axis momenta.
	This is clear from Figure~\ref{fig:cuts}, where the momentum cuts can be seen to represent the full range of available momenta.
	It can also be seen from the figure that although the cuts and on-axis momenta agree in the IR region, they start to diverge slightly in the UV. Since the UV behaviour of the spectral function can already be accessed perturbatively, the most interesting physics of the spectral function for our current research is given by the IR region. Using the momentum cuts therefore gives identical information to the on-axis momenta on the IR region while providing a better connection with the UV.

	\section{From dressing function to propagator}
	\label{app:rho_ghost}
	This Appendix will detail the relationship between the spectral density functions of the dressing function and the propagator.
	The dressing function is defined as
	\begin{equation}
		g(p) = p^2 G(p).
	\end{equation}
	Since ghost particles are massless, the spectral density function is expected to contain a $\delta$ function at zero momentum. Because such a $\delta$ peak complicates the numerical inversion, it is beneficial to remove it first by inverting the dressing function instead.
	Let
	\begin{align}
		g(p) = \int_{-\infty}^{\infty} \frac{\sigma(\omega)}{\omega - i p} \dd{\omega}\qquad \text{and} \qquad
		G(p) = \int_{-\infty}^{\infty} \frac{\rho(\omega)}{\omega - i p} \dd{\omega},
	\end{align}
	where both $\rho$ and $\sigma$ are odd functions. We then define
	\begin{equation}
		\hat{\rho} = - \frac{\sigma(\omega)}{\omega^2}.
	\end{equation}
	Note that the appearance of $\omega^2$ is motivated by the fact that both $\rho$ and $\sigma$ have to be odd functions.
	Integrating over $\hat{\rho}$ gives
	\begin{align*}
		\int_{-\infty}^{\infty} \frac{\hat{\rho}(\omega)}{\omega - i p} \dd{\omega} &= - \int_{-\infty}^{\infty} \frac{\sigma(\omega)}{\omega^2 \pqty{\omega - i p}} \dd{\omega} \\
		&= - \frac{i}{p} \cancel{ \int_{-\infty}^{\infty} \frac{\sigma(\omega)}{\omega^2} \dd{\omega} } - \frac{1}{p^2} \int_{-\infty}^{\infty} \frac{\sigma(\omega)}{\omega} \dd{\omega} + \frac{1}{p^2} \int_{-\infty}^{\infty} \frac{\sigma(\omega)}{\omega - ip} \dd{\omega} \\
		&= -\frac{g(0)}{p^2} + \frac{g(p)}{p^2} =  -\frac{g(0)}{p^2} + G(p),
	\end{align*}
	where the first term on the second line could be dropped due to the oddness of the integrand.
	We see that $\hat{\rho}$ nicely produces $G(p)$ but also an extra term. This extra term can be cancelled by identifying $\rho$ as
	\begin{equation}
		\rho(\omega) = - g(0) \delta'(\omega) + \hat{\rho}(\omega).
		\label{eq:rho_ansatz_repeat}
	\end{equation}

	\begin{figure}
		\begin{center}
			\vspace{-20pt}
			\includegraphics[width=0.8\textwidth]{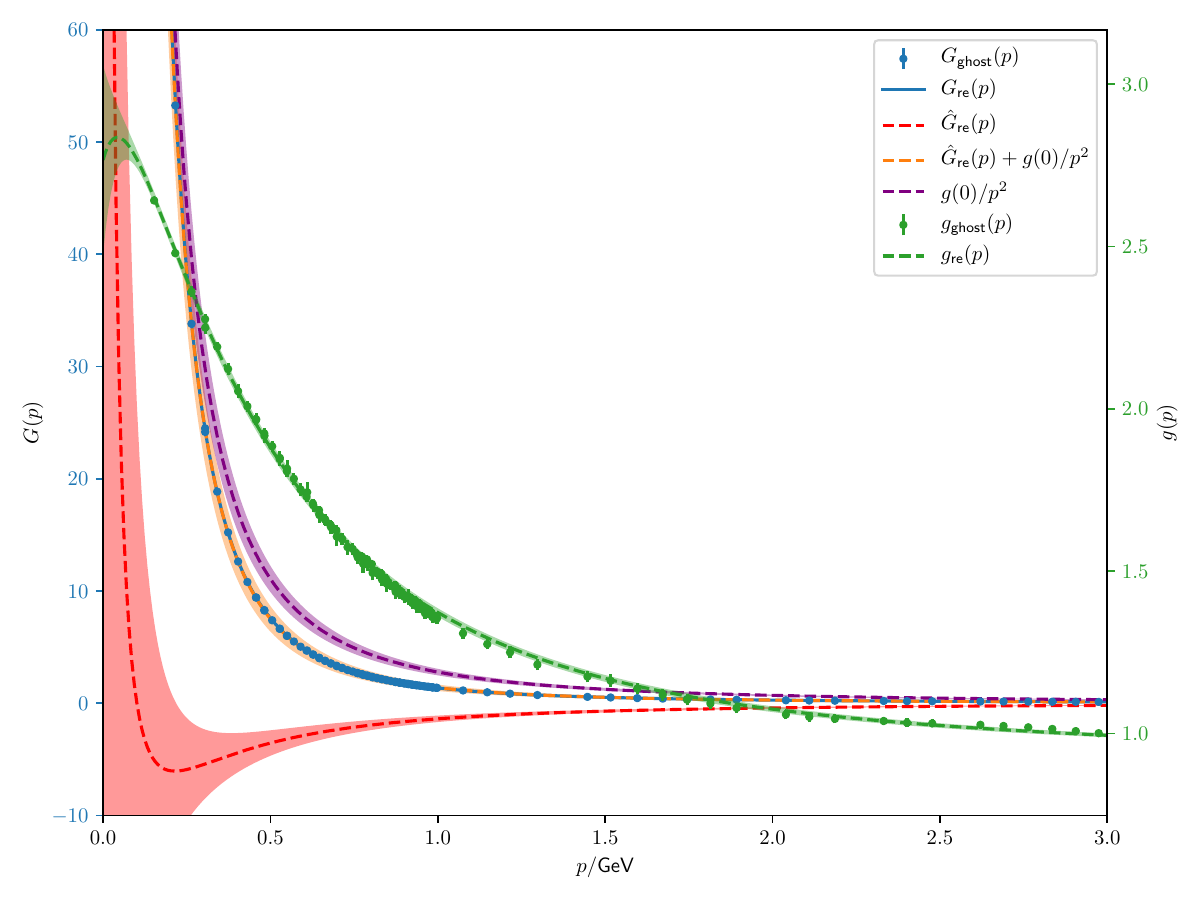}
			\caption{Reconstructions with a linear $G(p)$-axis.}
			\label{fig:G_ghost_lin}
		\end{center}
	\end{figure}
	We are forced to use the partial derivative operator $\delta'(\omega)$ instead of $\delta(\omega)$ due to the demand that $\rho(\omega)$ should be an odd function. Integrating over Eq.~\eqref{eq:rho_ansatz_repeat}, we find that his indeed gives $G(p)$.

	Figure \ref{fig:G_ghost_lin} shows this process numerically. As can be seen from the figure, $\hat{G}(p)$ can hardly be called a reconstruction of the data. However, upon adding $g(0)/p^2$ the result describes the data equally well as the direct reconstruction $G_\text{re}(p)$.

	\bibliographystyle{unsrt}
	\bibliography{biblio}

\end{document}